\documentclass[manuscript,screen]{acmart}

\renewcommand\footnotetextcopyrightpermission[1]{} % removes footnote with conference information in first column
\usepackage{multirow}

\usepackage{hyperref}

\usepackage[vlined,ruled,linesnumbered]{algorithm2e}
\makeatletter
\patchcmd{\@algocf@start}% <cmd>
  {-1.5em}% <search>
  {0pt}% <replace>
{}{}% <success><failure>
\makeatother

\usepackage{xcolor}
\usepackage{tikz}
\usetikzlibrary{fit,calc}
\newcommand*{\tikzmk}[1]{\tikz[remember picture,overlay,] \node (#1) {};\ignorespaces}
\newcommand{\boxit}[1]{\tikz[remember picture,overlay]{\node[yshift=3pt,fill=#1,opacity=.25,fit={(A)($(B)+(.94\linewidth,.8\baselineskip)$)}] {};}\ignorespaces}
\colorlet{gray}{gray!60}
\usepackage{siunitx}
\input{macros}
\AtBeginDocument{%
  \providecommand\BibTeX{{%
    \normalfont B\kern-0.5em{\scshape i\kern-0.25em b}\kern-0.8em\TeX}}}

\setcopyright{acmcopyright}
\copyrightyear{2021}
\acmYear{2021}
\acmDOI{XXXXXXX.XXXXXXX}

\begin{document}

\title{Disentangling Decentralized Finance (DeFi) Compositions}


% \author{Anonymous Author(s)\\}
% \email{anonymous@email.com}
% \affiliation{%
%   \institution{\phantom{institute}}
%   \city{\vspace{6em}}
%   \country{\phantom{country}}
% }

\author{Stefan Kitzler}
\email{kitzler@csh.ac.at}
\affiliation{%
  \institution{Complexity Science Hub Vienna and AIT - Austrian Institute of Technology}
  \city{Vienna}
  \country{Austria}
}
\author{Friedhelm Victor}
\email{friedhelm.victor@tu-berlin.de}
\affiliation{%
  \institution{Technische Universit{\"a}t Berlin}
  \city{Berlin}
  \country{Germany}
}
\author{Pietro Saggese}
\email{pietro.saggese@ait.ac.at}
\affiliation{%
  \institution{AIT - Austrian Institute of Technology and Complexity Science Hub Vienna}
  \city{Vienna}
  \country{Austria}
}
\author{Bernhard Haslhofer}
\email{haslhofer@csh.ac.at}
\affiliation{%
  \institution{Complexity Science Hub Vienna}
  \city{Vienna}
  \country{Austria}
}

\renewcommand{\shortauthors}{Kitzler et al.}

% !TeX root = ../main.tex
\begin{abstract}

% Background & Motivation

We present a measurement study on compositions of Decentralized Finance (DeFi) protocols, which aim to disrupt traditional finance and offer services on top of distributed ledgers, such as Ethereum. Understanding DeFi compositions is of great importance, as they may impact the development of ecosystem interoperability, are increasingly integrated with web technologies, and may introduce risks through complexity.
% Contributions & Findings
Starting from a dataset of \DeFiNumberProtocols labeled DeFi protocols and \DeFiNumberAccounts associated Ethereum accounts, we study the interactions of protocols and associated smart contracts. From a network perspective, we find that decentralized exchange (DEX) and lending protocol account nodes have high degree and centrality values, that interactions among protocol nodes primarily occur in a strongly connected component, and that known community detection methods cannot disentangle DeFi protocols.
Therefore, we propose an algorithm to decompose a protocol call into a nested set of building blocks that may be part of other DeFi protocols.
This allows us to untangle and study protocol compositions. With a ground truth dataset we have collected, we can demonstrate the algorithm's capability by finding that swaps are the most frequently used building blocks.
As building blocks can be nested, i.e., contained in each other, we provide visualizations of composition trees for deeper inspections. We also
present a broad picture of DeFi compositions by extracting and flattening the entire nested building block structure across multiple DeFi protocols. 
Finally, to demonstrate the practicality of our approach, we present a case study that is inspired by the recent collapse of the UST stablecoin in the Terra ecosystem. Under the hypothetical assumption that the stablecoin USD Tether would experience a similar fate, we study which building blocks and, thereby, DeFi protocols would be affected.
Overall, our results and methods contribute to a better understanding of a new family of financial products.
\end{abstract}



\maketitle

% !TeX root = ../main.tex

\section{Introduction}\label{sec:intro}

% Background

Decentralized Finance (DeFi) stands for a new paradigm that aims to disrupt established financial markets. It offers financial services in the form of \emph{smart contracts}, which are executable software programs deployed on top of distributed ledger technologies (DLT) such as Ethereum. Despite being a relatively recent development, we can already observe rapid growth in DeFi protocols enabling lending of virtual assets, exchanging them for other virtual assets without intermediaries, or betting on future price developments in the form of derivatives like options and futures. The term ``financial lego'' is sometimes used because DeFi services can be \emph{composed} into new financial products and services.

\begin{figure}
	\centering
	  \vspace{1em}
    \includegraphics[width=1\linewidth]{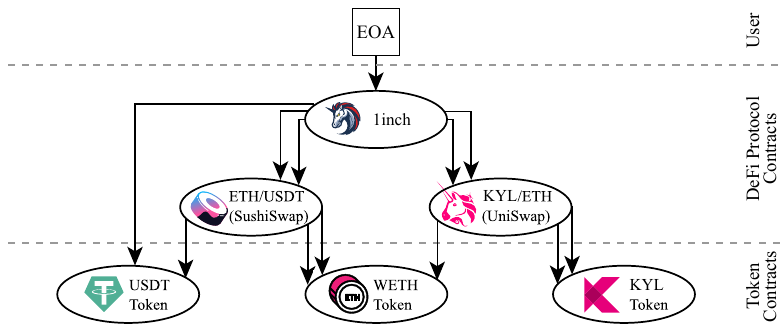}
		\vspace{0.5em}
	\caption{A DeFi composition where \textit{USDT} tokens are swapped against \textit{KYL} tokens through the DeFi service \emph{1inch} in a single transaction. \emph{1inch} executes the swap sequentially through the DeFi services \emph{SushiSwap} and \emph{UniSwap}, using \textit{WETH} as an intermediary token. In the transaction trace graph, we can see the user calling the \emph{1inch} smart contract, which in turn triggers several calls to DeFi protocol-, and token smart contracts.}
	\label{fig:composition_example}
	\Description{A DeFi composition where \textit{USDT} tokens are swapped against \textit{KYL} tokens through the DeFi service \emph{1inch} in a single transaction. \emph{1inch} executes the swap sequentially through the DeFi services \emph{SushiSwap} and \emph{UniSwap}, using \textit{WETH} as an intermediary token. In the transaction trace graph, we can see the user calling the \emph{1inch} smart contract, which in turn triggers several calls to DeFi protocol-, and token smart contracts.}
\end{figure}

As an example of a DeFi composition, consider Figure~\ref{fig:composition_example}, which illustrates a user interacting with the \emph{1inch} decentralized exchange (DEX) aggregator Web service\footnote{\url{https://app.1inch.io} (this and all the following links were accessed on June 15 2022)}. The user holds an amount of \textit{USDT} tokens and wants to swap them to \textit{KYL} tokens. Using the Web application and her \emph{externally owned account (EOA)}, she creates a transaction against the \emph{1inch} contract, which in turn triggers a sequence of two swaps on two DeFi protocols within the same transaction, from \textit{USDT} to \textit{WETH} on \emph{SushiSwap} and thereafter from \textit{WETH} to \textit{KYL} on \emph{UniSwap}.
In this paper, we study such single transaction DeFi interactions and the networks that arise when combining multiple DeFi transactions.

\subsection{Motivation}

In \num{2021}, the total value of tokens held by smart contracts underlying the DeFi protocols has reached \DeFiTotalValue billion USD~\cite{defipulse:2021a}, demonstrating rapid growth. As composability of DeFi protocols is frequently seen as one of the main advantages~(cf.~\cite{Schaer:2021a}), there are multiple reasons why it is interesting to study DeFi compositions:

\noindent\textbf{Ecosystem interoperability}. While composability can be seen as an opportunity, single transaction compositions as shown in Figure~\ref{fig:composition_example} currently only work within a single distributed ledger.
Most of the emerging DLT scaling solutions, such as sidechains~\cite{singh2020sidechain, Mohanty2022BlockchainInteroperability}, rollups~\cite{thibault2022blockchain}, and off-chain networks~\cite{gudgeon2020sok, sguanci2021layer}, lead to multiple, somewhat isolated DeFi ecosystems. Hence, composability is disrupted, as smart contracts on one platform cannot invoke contract functions on another platform within a single transaction. Understanding which types of compositions are frequently used may help in developing solutions to cross-chain~\cite{zamyatin2021sok, belchior2021survey, Herlihy2018AtomicCrossChain, wang2021sok} composability.
Until solutions are found, such knowledge can help in deciding which services should be co-located, and which services could be separate.

\noindent\textbf{Integration with Web technologies}. Cryptoassets have started integrating with various Web technologies. For example, the Brave\footnote{\url{https://brave.com}} browser includes an integrated cryptoasset wallet and native use of BAT tokens, and various applications from the commercial BitTorrent\footnote{\url{https://www.bittorrent.com}} ecosystem rely on the BitTorrent Token (BTT). This raises the question regarding the interdependence between DeFi compositions and web technologies. Services like Furucombo\footnote{\url{https://furucombo.app}} already illustrate that almost arbitrary DeFi compositions are constructed through Web interfaces. In order to develop an understanding of this, however, it is important to identify compositions and their points of interaction in the first place.

\noindent\textbf{Risks through complexity.} After its deregulation in the early 2000s, the securitization market became more complex and opaque. Financial institutions used new financial instruments to maximize their exposure in this market. They were based on technical computer models and traded by highly leveraged institutions, many of whom did not understand the underlying models. These instruments were highly profitable, but the lack of any infrastructure and public information about them created a massive panic in the financial system that began in August 2007~\cite{Baily:2008a}.
DeFi protocols may offer opportunities, such as technological innovation or new governance models. However, their composability adds additional complexity and opaqueness to an already complex cryptoasset ecosystem, which currently has a market valuation of about 1T USD\footnote{\url{https://coinmarketcap.com/charts/}}. If these protocols are not understood and adopted more broadly, they could have unforeseeable systemic effects on financial markets and our society as a whole, as seen in the 2008 financial crisis~\cite{kirilenko2017flash}.
A recent example involving DeFi protocols is the collapse of the stablecoin protocol Terra and its associated cryptoassets LUNA and UST. While the protocol did work as designed, its stabilization mechanism was not robust to significant selling pressure in the advent of market participants panicking. This ultimately led to deleveraging spiral effects~\cite{klages2021In}  destroying over 30B USD of value within a single week and rendering institutions with large exposures to LUNA or UST insolvent. In addition, the stablecoin UST was used as part of compositions in many other DeFi protocols on the Terra blockchain and through bridges on different blockchains, thus affecting the entire ecosystem~\cite{Rai:2022a}.

Previous work (cf.,~\cite{Daian:2020aa,Gudgeon:2020aa}) has partially studied risks in the DeFi ecosystem, showing possible strategies that allow rational agents to maximize their revenues by subverting the intended design of DeFi protocols,
for example in DEXs and lending protocols.
However, none of the existing studies have systematically investigated compositions of DeFi protocols, which form complex, interconnected financial instruments. 

\subsection{Contributions}

Our work aims to analyze DeFi protocols and to develop a novel algorithmic method that helps to understand protocol compositions. We can summarize our contributions as follows:

\begin{enumerate}

	\item We provide a manually curated ground truth of \DeFiNumberSeedAddr addresses from \DeFiNumberProtocols DeFi protocols and derived \DeFiNumberAccounts associated Ethereum smart contracts. These are labels that can be reused in future research. On this basis, we propose two network abstractions, representing interactions among DeFi protocols and smart contracts (Section~\ref{sec:data_methods}).

	\item We study intertwined DeFi protocols from a macroscopic perspective by analyzing the topology of both networks. We find that DEX and lending protocols have high degree and centrality values, and protocol interactions primarily occur in a strongly connected component. We also find that known community detection algorithms can only indicate DeFi compositions but cannot effectively disentangle them (Section~\ref{sec:topology}).

	\item We address the microscopic transaction level and propose an algorithm for extracting the building blocks of DeFi protocols. We apply the algorithm to all protocol transactions in our ground truth, identify the most frequent building blocks, and find that swaps are the most frequent ones. We show how the observed space of compositions looks like for the \emph{Aave} protocol. Further, we also demonstrate, using \emph{1inch} and \emph{Instadapp} as examples, how to disentangle and visualize the building blocks of a single protocol as a treemap (Section~\ref{sec:buildingBlocks}).

	\item We present an overall picture of DeFi compositions by extracting and flattening the entire nested building block structure across multiple DeFi protocols. The results show that DeFi aggregation protocols (\emph{1inch}, \emph{0x} or \emph{Instadapp}) are, as expected, heavily intertwined with many other DeFi protocols, which confirms that our algorithm works as intended (Section~\ref{sec:block_analysis}).

	\item Finally, we present a case study illustrating how a hypothetical run on the stablecoin USD Tether would affect the building blocks of individual DeFi protocols. (Section~\ref{sec:case_study}). We detect a comparatively high dependency of \emph{Curvefinance} building blocks to the USDT cryptoasset.

\end{enumerate}

\noindent We believe that our results are an essential contribution towards understanding DeFi compositions. On a microscopic level, our proposed methods can be used to assess the composition of individual protocols. On a macroscopic level, they show how DeFi protocols and their implementations are connected with each other. For this paper, we limit our scope to the largest Ethereum Virtual Machine (EVM)-based blockchain Ethereum, but in principle the approach can be used and applied to any other EVM-based platform.
For reproducibility of results, we make our ground truth dataset, including the labels as well as our source code, openly available at \url{https://github.com/StefanKit/Untangling_DeFi_Composition}.

% !TeX root = ../main.tex

\section{Background and Definitions}\label{sec:background}

We now establish preliminary terms and definitions that are used throughout this work and introduce the related works.

\subsection{Ethereum Account Types}

Ethereum is currently the most important distributed ledger technology (blockchain) for DeFi services~\cite{Zetzsche:2020aa}. It differs from the Bitcoin blockchain conceptually as it implements the so-called ``account model'' with two different account types.
An \textit{externally owned account} ($EOA$) is a ``regular'' account controlled by a private key held by some user. A \textit{code account} ($CA$), which is synonymous with the notion ``smart contract'', is an account controlled by a computer program, which is invoked by issuing a transaction with the code account as the recipient.

A \CA must always be initially called by an \textit{external transaction} originating from an $EOA$, but a \CA can itself trigger other \CAs.
In the latter case, the interaction, which is also known as ``message'', is denoted as an \textit{internal transaction}.
Several branches of internal transactions with varying depth can follow an external transaction, resulting in cascades, which altogether are called \textit{traces}.

\CAs allow users to implement application-layer protocols, which are essentially programs that can follow some standardized interface. \textit{Tokens} are popular \CA-based applications and a way to define arbitrary assets that can be transferred between accounts. The program behind a token manages token ownership and can implement a standardized interface like ERC20, which defines functions standardizing token transfer semantics.

\subsection{Decentralized Finance (DeFi) Protocol}

A \textit{DeFi protocol} is an application-layer program that provides financial service functions such as swapping or lending assets. More technically, we can define it as follows:

\begin{definition}
    A DeFi protocol $P$ is a decentralized application that facilitates specific financial service functions defined and implemented by a set of protocol-specific code accounts.
\end{definition}

The following properties distinguish DeFi services from traditional financial services: first, they are \emph{non-custodial}, meaning that no intermediary such as a bank or a broker holds custody of users' funds. Second, they are \emph{permissionless}, meaning that anyone can use existing or implement new services. Third, they are \emph{transparent}, which means that anyone with the necessary technical capabilities and skills can investigate and audit the state of protocols. The fourth is that DeFi protocols are \emph{composable}.

\subsection{DeFi Protocol Compositions}

The last property, \emph{composability}, is the most crucial for this work and requires more detailed description: \CAs can call each other, and their individual functions can be arbitrarily composed into new financial products and services (``Financial Lego'')~\cite{Werner:2021to}.
While this analogy is widely used in the literature, to the best of our knowledge, no work investigates \emph{which are} the basic composable building blocks of more complex financial services and how they are related.
Harvey et al.~\cite{harvey2021defi} refer broadly to composability as asset tokenization and networked liquidity, while Von Wachter et al.~\cite{wachter2021measuring} conceive composability narrowly as a repeated wrapping operation of tokens resulting in new derivative products. However, as illustrated before in Figure~\ref{fig:composition_example}, we note that DeFi compositions also involve \CAs, which are not tokens.
Also, Engel and Herlihy~\cite{engel2021composing} and Tolmach et al.~\cite{tolmach2021formal} respectively discuss compositions only in the context of automated market makers (AMMs) and of formal verification of \CAs related to decentralized exchanges and lending services, which is again a very narrow conception.
Thus, there is no comprehensive, technically grounded definition for DeFi compositions to the best of our knowledge. For our work, we define it as follows:

\begin{definition}
    A DeFi protocol composition occurs when a protocol-specific account leverages, within a single transaction, one or more accounts belonging to the same or another DeFi protocol to provide a novel financial service.
\end{definition}

\subsection{Related Work}
\label{sec:litrev}

Others studied networks closely related to the ones we investigated before us:
Guo et al.~\cite{guo2019graph} are amongst the first to investigate the Ethereum transaction graph, finding that volumes moved and the numbers of transactions follow a power law distribution, that the component structure follows a bow-tie model, and that negative assortativity is plausibly explained by the presence of service providers such as exchanges.
Chen et al.~\cite{chen2020understanding} conduct a systematic study of Ethereum between 2015 and 2018 and exploit graph analysis measures to describe three different network constructions (money transfer, smart contract creation, and smart contract invocation).
Another systematic study has been conducted by Lee et al.~\cite{Lee:2020wa}, who analyzed the local and global properties of interaction networks extracted from the entire Ethereum blockchain statically finding heavy-tailed degree distributions. In a follow-up, Zhao et al.~\cite{Zhao:2021uk} analyzed the temporal evolution of Ethereum interaction networks and found that they proliferate and follow the preferential attachment growth model.
Furthermore, several studies focus on the network of Ethereum's tokenized assets: Somin et al.~\cite{Somin:2018aa}, for instance, studied the combined graph of all fungible token networks, while Victor and L{\"u}ders~\cite{Victor:2019wt} explored the networks of the top 1,000 ERC20 tokens individually. Fr{\"o}wis et al.~\cite{Frowis:2019uv} proposed a method for detecting token systems independent of an implementation standard.
Also, Chen et al.~\cite{Chen:2020uo} conducted a systematic investigation of the whole Ethereum ERC20 token ecosystem and analyzed their activeness, purpose, relationship, and role in token trading.
Other studies exploited network methods for the detection of specific nodes using graph-based approaches. Poursafaei et al.~\cite{poursafaei2021sigtran} developed a method based on graph node feature extraction and graph representation learning techniques to identify illicit nodes. Li et al.~\cite{li2020dissecting} and Ofori-Boateng et al.~\cite{ofori2021topological}, instead, respectively use Topological Data Analysis (TDA) to detect price anomalies and hidden co-movement in pairs of tokens, and for anomalous events
detection in  a multilayer network.
However, none of these related works considers networks that represent DeFi Protocols and their relationships.

Another growing body of research concentrates on specific functions offered by individual DeFi protocols or types of protocols.
We are aware of many DEX-related measurements focusing on protocol-specific aspects, such as the magnitude of cyclic arbitrage activity~\cite{wang2021cyclic}, the behavior of liquidity providers~\cite{wang2021behavior}, or the role of oracles as providers of external information~\cite{liu2020first}.
Other studies focus on lending and borrowing services: Perez et al.~\cite{perez2020liquidations} analyze liquidations and related participants' behavior in the DeFi protocol \emph{Compound}, while Gudgeon et al.~\cite{gudgeon2020defi} compare market efficiency, utilization, and borrowing rates in different lending protocols. Also, Wang et al.~\cite{wang2021towards} provide methods to identify flash loans in three different DeFi providers and measure their related activity.
Finally, we are aware that von Wachter et al.~\cite{wachter2021measuring} investigate composability from an asset perspective and measure composability by identifying the number of derivatives produced from an initial root asset. However, we apply a more technical, service-oriented perspective and consider, to put it simply, a DeFi composition as being a computer program utilizing other programs' functions.

Overall, we are not aware of previous studies providing a comprehensive picture of DeFi compositions across various protocols. We also do not know any work that analyzes in detail the building blocks of individual DeFi protocols. With this work, we want to close this gap.

% !TeX root = ../main.tex

\section{Dataset and Network Construction}\label{sec:data_methods}

This section describes the data we collected and the network abstractions we constructed for subsequent analysis steps.

\subsection{Dataset collection}

To study DeFi compositions, we are interested in transactions between Ethereum code accounts associated with known DeFi protocols. Thus, we used on-chain transaction data from the Ethereum blockchain and built a ground truth of known \CAs and their associations to DeFi protocols.

\subsubsection{On-chain transaction data}
While Ethereum's history goes as far back as July 2015, DeFi only emerged as a popular term around summer 2020, when these protocols first saw increased usage. This informed our choice of the analysis time frame and the ability to refer to external sources providing information on popular, established DeFi services. 
We used an OpenEthereum client and ethereum-etl\footnote{\url{https://github.com/blockchain-etl/ethereum-etl}} to gather all Ethereum transactions from \TracesFrom to \TracesTo. %
We collected each external transaction and also parsed its cascade of internal transactions, which together give us the \emph{trace}. For each transaction, we extracted the source and destination account addresses, the transaction hash, the transferred value, the transaction type (call, create, or self-destroy), as well as the trace ID, which indexes the transactions by their execution order. Additionally, we collected the method ID of the 4-byte input sequence, which allows us to identify the signature of called methods using the 4Byte lookup service\footnote{\url{https://www.4byte.directory/}}.

To distinguish between \CAs and \EOAs, we gathered all code account creation transactions from the first \CA created on Ethereum until the end of our observation period. We also use these \creationTraces to associate each \CA with its creator \CA. In total, we found \contractsNumber \CAs and used the output byte sequence to identify \ERCNumber contracts conforming to the ERC20 standard.

\subsubsection{Ground truth data}\label{sec:ground_truth_data}
To be able to analyze DeFi protocols, we need a ground truth dataset on which smart contracts are part of a given protocol.
We focus on the most relevant protocols regarding valuation and gas-burned between \SeedFrom and \SeedTo, using monthly samples of the top three total-value-locked (TVL) protocols from DeFi Pulse\footnote{ \url{https://defipulse.com/}} for each financial service category.
Additionally, we consider protocols including \CAs of the top ten gas burner list\footnote{\url{https://ethgasstation.info/gas-burners}} in the observation period. The result defines the set of DeFi protocols we want to investigate.
Table~\ref{tab:sum_seed} reports summary statistics for the 23 protocols in our sample, divided by category.
The last column reports, for each protocol, the average share of each protocol's TVL with respect to the entire DeFi ecosystem, between March and August 2021. 
In total, our 23 DeFi protocols cover more than 81\% of the entire DeFi TVL. According to DeFi Pulse, in August 2021 more than a hundred DeFi protocols existed, but only around 30 (of which 18 in our sample) had a TVL larger than 200M USD.
Most of the protocols in our sample are still the most relevant ones for TVL as of June 2022\footnote{10 out of the first 11 DeFi protocols for TVL in DeFi Pulse are in our dataset.}.
In the following, we briefly introduce the categories and protocols as reported by DeFi Pulse\footnote{DeFiPulse reports the protocols divided into five categories. We don't include the \textit{Payment} category because services like Polygon provide off-chain functionality rather than composable financial services or products.}:
\begin{itemize}
    \item \textbf{Assets} identify the category including cryptoasset management protocols, such as yield aggregators, that aim at maximizing the value of a portfolio or basket of underlying assets.
    \textit{Harvestfinance}, \textit{Yearn}, \textit{Vesper}, share a similar mechanism, whereby they pool resources which are in turn invested in other DeFi platforms according to different optimization strategies. Users are typically rewarded through tokenized assets. 
    \textit{Convex} enables Curvefinance liquidity providers to earn additional rewards. 
    \textit{Badger} allows Bitcoin users to deposit tokenized Bitcoin such as wBTC and consequently generate a yield, by following programmatic optimization strategies. Similarly, \textit{RenVM} bridges digital assets across DeFi ecosystems by minting ERC20 tokens on Ethereum with 1:1 ratio.
    \textit{Fei}'s protocol builds on a decentralized stablecoin backed by cryptoassets exploited through yield strategies established by the protocol's governance.
    \item \textbf{Derivatives} protocols allow issuing synthetic financial instruments in the DeFi ecosystem, either tracking other cryptoassets or real-world off-chain assets. 
    \textit{Synthetix}, for instance, supports several real-world assets, such as fiat currencies and metals, while \textit{dYdX} allows investors to trade perpetual positions on the underlying cryptoassets. \textit{Hegic} enables the issuing of ETH and wBTC call and puts options. \textit{Futureswap} users can open leveraged long and short positions on cryptoassets.
    \textit{Nexus}, instead, provides financial insurance instruments that cover potential losses users might incur in; similarly, \textit{Barnbridge} offers tools to hedge risk through its financial instruments.
    \item \textbf{DEXs}, i.e. Decentralized Exchanges, allow users to exchange cryptoassets. \textit{UniSwap}, \textit{SushiSwap}, \textit{Curvefinance}, \textit{Balancer} all exploit Automated Market Makers (AMM), as well as bonding curves and constant functions to algorithmically set the cryptoassets prices, while \textit{0x} is based on the order book mechanism. The \textit{1inch} protocol aggregates information on liquidity from several DEXs and routes transactions to those offering the best prices.
    \item \textbf{Lending} protocols provide investors with automated markets for loanable funds: lenders issue interest-bearing instruments and borrowers can take positions, typically conditional to the provision of collateral that covers potential losses. \textit{Aave} and \textit{Compound} follow the model described above.
    \textit{Maker} users lock their cryptoassets as collateral and receive the DAI token in return.
    \textit{Instadapp} follows a more complex scheme and acts mostly as an aggregator of multiple DeFi protocols.
\end{itemize}

After identifying the most relevant DeFi protocols, we manually collected the \CAs associated with each protocol. Since this information is not available on the blockchain, we rely on off-chain and publicly available sources like protocol websites and available documentation. We resolved conflicts of duplicated \CA to protocol assignments and identical names by querying \CA addresses on Etherscan\footnote{\url{https://etherscan.io/}} and uniquely assigned each \CA address to its original protocol and obtained a unique label. We denote these manually collected data points as \emph{seed data} and make them available as part of our source code repository.

Next, we extended our seed data by implementing a heuristic that uses the creation transactions and identifies the \CAs deployed by each seed address. By default, all extended addresses inherit the label and protocol assignments from the corresponding seed address. If the procedure leads to a conflict of labels for an address, we preserve the one obtained through the heuristic. Combined with our seed data, these extended addresses form our \emph{extended seed data} set. 
Table~\ref{tab:sum_seed} summarizes the number of seed and extended addresses collected for each DeFi protocol. It shows that our automated expansion does not increase the number of addresses associated with DeFi protocols for assets and derivatives. However, it massively expands the dataset for DEXs and lending protocols utilizing automated factory contract deployments. In particular, more than 10 million additional \CAs are associated with \emph{1inch} due to the factory contract that deploys gas tokens.
The last column shows the number of external transactions directed to each of our DeFi protocols. The distribution is heterogeneous, and again the most relevant categories are DEX and lending. \emph{UniSwap} is the most frequently appearing one, with a gap of around one order of magnitude to the second one, which is \emph{Maker}.

\begin{table}
	\centering
	\caption{Ground truth dataset summary statistics. Seed addresses were manually collected for each DeFi protocol. The extended seed are heuristically derived and include also further created code accounts from the seed addresses.}
		\begin{tabular*}{\columnwidth}{@{\extracolsep{\fill}}lrrrrr}%\begin{tabular}{>{\centering\arraybackslash}p{0.1\textwidth}>{\centering\arraybackslash}m{0.1\textwidth}>{\centering\arraybackslash}m{0.07\textwidth}>{\centering\arraybackslash}p{0.13\textwidth}}
\toprule  &  & \multicolumn{2}{c}{Number of addresses} & & \\ \cmidrule{3-4} 
Protocol type &       \makecell[c]{DeFi \\ Protocol} &  Seed &  Extended seed  & External calls & \% TVL \\
\midrule
     \multirow{7}{*}{Assets} &         Badger &    \num{64} &            \num{278}  & \num{258773}  & 1.09\% \\
      &         Convex &    \num{22} &   \num{131}  &  \num{147855}  &  1.13\% \\
      &            Fei &    \num{40} &             \num{37}  &  \num{146691} & 0.28\% \\
      & Harvestfinance &   \num{101} &            \num{803}  &  \num{119631}  &  0.46\%  \\
      &          RenVM &    \num{15} &             \num{15}  &  \num{234161}  & 0.86\%  \\
      &         Vesper &    \num{44} &             \num{44}  &  \num{94189}  & 1.19\% \\
      &          Yearn &     \num{3} &              \num{3}  &  \num{243036}  &  3.54\% \\
               \midrule%\specialrule{.05em}{0.3em}{0.3em} 
\multirow{6}{*}{Derivatives} &     Barnbridge &    \num{40} &             \num{46}  &  \num{55588} &  0.17\% \\
 &           dYdX &    \num{38} &             \num{38}  & \num{107264}  & 0.14\%  \\
 &     Futureswap &     \num{9} &             \num{10}  & \num{6484}  & 0.04\%  \\
 &          Hegic &     \num{8} &              \num{8}  & \num{8372}  & 0.03\%   \\
 &          Nexus &    \num{24} &             \num{26}  & \num{20067}  &  0.57\%  \\
 &      Synthetix &   \num{271} &            \num{272}  & \num{611942}  & 2.55\%  \\
          \midrule
        \multirow{6}{*}{DEX} &             0x &    \num{28} &             \num{50}  & \num{2094335}  & - \%  \\
         &          1inch &    \num{15} &       \num{10338305}  & \num{1277641}  & 0.52\%   \\
         &       Balancer &     \num{9} &           \num{3473}  & \num{281530}  & 2.29\%  \\
         &   Curvefinance &   \num{163} &            \num{267}  & \num{745672}  & 9.28\%  \\
         &      SushiSwap &    \num{12} &           \num{1705}  & \num{2026674}  & 5.37\%  \\
         &        UniSwap &    \num{15} &          \num{54038}  & \num{28394798}  & 8.30\%  \\
         \midrule 
    \multirow{4}{*}{Lending} &           Aave &   \num{157} &            \num{166}  &  \num{851578}  & 13.31\%  \\
     &       Compound &    \num{67} &             \num{65}  &  \num{741069}  & 11.48\%  \\
     &      Instadapp &    \num{72} &          \num{32770}  &  \num{97080}  & 7.39\%  \\
     &          Maker &   \num{190} &         \num{231261}  &  \num{2992692}  & 11.77\%  \\
\bottomrule
\end{tabular*}

	\label{tab:sum_seed}
\end{table}

\subsubsection{Dataset reduction}

As we are only interested in known DeFi protocols, we finally limited and reduced the \traces data set to the subset of \protocolTraces, where the initial external transaction originating from an \EOA triggers a \CA address in our extended seed dataset. This reduction allows us to investigate and interpret compositions within the context of known protocols.

\subsection{Network construction}

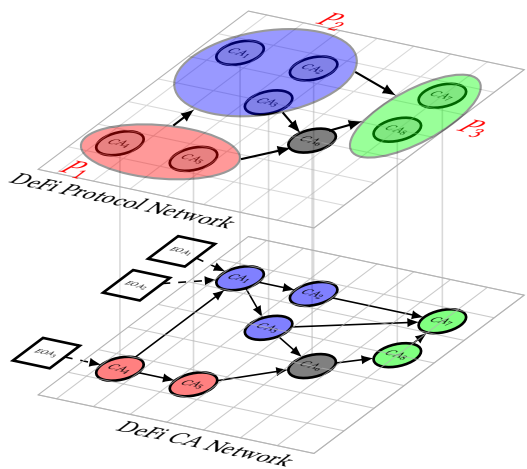
\begin{figure}
	\centering
	\begin{tikzpicture}[multilayer=3d,scale = 1]
\SetLayerDistance{-3}
\Vertices{figures/network_ill_vertices1.csv}
\Vertices{figures/network_ill_vertices2.csv}
\Edges{figures/network_ill_edges12.csv}
\Edges{figures/network_ill_edges2.csv}
\begin{Layer}[layer=1]
\draw[gray] (0.75,0) rectangle (4.5,3.5);
\draw[step=.5, gray,draw opacity=.5] (0.75,0) grid (4.5,3.5);
\node at (0.5,0)[below right,black]{DeFi Protocol Network};

% P_1
\draw (1.8,0.75) node[ellipse, minimum width=2cm, minimum height=1cm, draw,thick, fill = red, opacity = 0.4] (ell1) {} node[color = red, ultra thick, anchor = north east, outer sep = 10pt,font= \Large] {$P_{1}$};
%
%P_2
\draw (1.8,2.5) node[ellipse, minimum width=2cm, minimum height=1.8cm, draw,thick, fill = blue, opacity = 0.4] (ell2) {} node[color = red, ultra thick, anchor = south, outer sep = 25pt,font= \Large] {$P_{2}$};
%
%P_3
\draw (4,2.3) node[ellipse, minimum width=1cm, minimum height=1.8cm, draw,thick, fill = green, opacity = 0.4] (ell3) {} node[color = red, ultra thick, anchor = west, outer sep = 15pt,font= \Large] {$P_{3}$};
%
%Unkn - needed to draw arrows
\draw (3.25, 1.5) node[circle,draw= none, fill = white,scale = 1.5,opacity = 0] (circ) {};

\draw[-{latex[scale=3.0]},thick] (ell1.115) -- (ell2.255);
\draw[-{latex[scale=3.0]},thick] (ell2.20) -- (ell3.150);
\draw[-{latex[scale=3.0]},thick] (ell2.320) -- (circ.160);
\draw[-{latex[scale=3.0]},thick] (circ.80) -- (ell3.210);
\draw[-{latex[scale=3.0]},thick] (ell1.10) -- (circ.240);

%\fill[rotate=0,thick,color = orange,opacity = 0.4] (1.8,0.75) ellipse (1cm and 0.5cm) node[color = red, ultra thick, anchor = north east, outer sep = 10pt,font= \Large] {$P_{1}$};
%\fill[thick,color = orange,opacity = 0.4] (1.8,2.5) ellipse (1.cm and 0.9cm) node[color = red, ultra thick, anchor = south, outer sep = 30pt,font= \Large] {$P_{2}$};
%\fill[thick,color = orange,opacity = 0.4] (4.,2.3) ellipse (0.5cm and 0.9cm) node[color = red, ultra thick, anchor = west, outer sep = 20pt,font= \Large] {$P_{3}$};
\end{Layer}
\begin{Layer}[layer=2]
%\filldraw[rotate=8,thick,color = gray,opacity = 0.6] (1.4,0.1) ellipse (1.2cm and 0.5cm);
\draw[gray] (0.75,0) rectangle (4.5,3.5);%(4.5,3);
\draw[step=.5, gray,draw opacity=.7] (0.75,0) grid (4.5,3.5);%(4.5,3);
\node at (4.5,0)[below left,black]{DeFi CA Network};
\end{Layer}
\end{tikzpicture}
	\caption{Schematic illustration of constructed networks. The lower-level DeFi Code Account (CA) network represents interactions between \CAs. The higher-level DeFi Protocol Network models relations between DeFi protocols. Lower-level \CA vertices are associated with higher-level protocol vertices. \CAs are triggered by \EOAs or other \CAs.}
	\label{fig:network_construction}
\end{figure}

In our analysis, we want to understand and discover relations between DeFi protocols and associated \CAs. For that purpose, as shown in Figure~\ref{fig:network_construction}, we constructed networks consisting of DeFi traces on two abstraction levels: the lower-level \emph{DeFi Code Account (CA) Network} and the higher-level \emph{DeFi Protocol Network}.

The DeFi \CA Network includes all known ground truth \CAs triggered by external transactions from arbitrary \EOA addresses and all \CAs subsequently called by cascades of internal transactions. We note that \CAs in the network can or cannot be associated with a DeFi protocol in our ground truth dataset. We construct the network by filtering all internal and external transactions between \CAs from the \protocolTraces. Since repeated usage of DeFi services results in recurring transaction patterns, we aggregate and count transactions with the same source and destination address.

The DeFi Protocol Network represents interactions between protocols. We constructed it by merging all DeFi \CA vertices associated with the same DeFi protocol into a single node.
We note that we modeled both networks as a directed graph, in which vertices represent either a protocol or a single \CA. The weighted edges represent the aggregated set of transactions between DeFi protocols or \CAs.

% !TeX root = ../main.tex

\section{Topology measurements}\label{sec:topology}

\begin{table}
    \centering
    \caption{Summary statistics of the analyzed networks.}
    \begin{tabular*}{\columnwidth}{@{\extracolsep{\fill}}lrr}
%\begin{tabular}{lcc}
    \toprule
    & DeFi CA network & DeFi Protocol network \\
    \midrule
    Nodes & \num{2536371} & \num{43624} \\
    Edges & \num{3472757} & \num{84789} \\
    Self-loops & \num{6668} & 146 \\
    Average degree & 1.369 & 1.944 \\
    Density & 5.398e-07 & 4.456e-05 \\
    %\textcolor{violet}{Reciprocity} & 0.259 & 0.215 \\
    %\textcolor{violet}{Assortativity} & -0.473 & -0.262 \\
    \bottomrule
\end{tabular*}
    \label{tab:sum_metrics}
\end{table}

We now analyze the constructed networks from a macroscopic perspective.
Since our research focuses on understanding DeFi compositions, we do not aim at conducting an encompassing study of the entire Ethereum topology, as it was done in previous studies (see Section~\ref{sec:litrev}). 
This supports our choice to focus on a narrower number of targeted metrics that provide relevant insights on composability aspects; other approaches that are beyond the scope of our work are discussed in Section~\ref{sec:limit}. 
The analysis of the degree distribution and centrality measures can help identifying the CAs implementing core functionalities, and the reciprocity and assortativity provide additional insights on the relationships across such CAs. 
To understand how CAs associated to the same protocols interact with each other, we investigate how the network is separated in different components and whether known community detection algorithms identify community structures that overlap or not with the protocols structures.

We start by reporting basic summary statistics for the DeFi CA network and the DeFi Protocol network in Table~\ref{tab:sum_metrics}. The main difference is in the network dimension, the latter being two orders of magnitude smaller. The presence of self-loops indicates that some contracts include multiple functionalities and thus can also call themselves. Both networks are sparse, as shown by the average degree and density measure, suggesting that \CAs tend to interact with only a few other \CAs.

\subsection{Degree distribution}
\label{sub:powerlaw}

\begin{figure}[b]
	\input{./figures/ccdf_together.tex}
	\caption{Degree distribution of the \CA ({\protect\tikz \protect\draw (0,0) node[cross=3pt,black!50!green,thick]{};}) and Protocol ({\protect\tikz \protect\fill[black!50!blue] (1ex,1ex) circle (0.5ex);}) networks are shown in the plot as cumulative distribution function (CCDF). The estimated parameters $\hat \theta  = (\hat k_{min}, \hat \alpha)$ are respectively $\hat \theta_{CA}  = (93, 1.69)$ and $\hat \theta_{P}  = (25, 1.83)$. In both networks, high-degree nodes are associated to DEX or lending protocols. For the \CA network, they are routing contracts or factory contracts that deploy other contracts. Nodes with high degree are likely to contain core functionalities and thus to play a relevant role in compositions.}
	\label{plot:deg_dist}
\end{figure}

Looking at the total-value-locked at DeFi Pulse, we can observe that some DeFi protocols and their contracts play a major role. This observation suggests that they might implement core functionality, which other protocols in DeFi compositions can in turn utilize. Under this assumption, preferential attachment~\cite{price1976general,barabasi1999emergence} is a plausible generative mechanism for both networks. More generally, networks whose degree distribution follows a power law, i.e., the fraction of vertices with degree $k$ is given by $ P(k) \sim  k^{-\alpha} $ for values of $k \geq k_{min}$, are often associated to such generative mechanism. We thus estimate the parameters $\hat \theta  = (\hat k_{min}, \hat \alpha)$ for our two networks and investigate if the power law distribution is a good fit.

% Method

We rely on the methodology introduced by Clauset et al.~\cite{clauset2009power} and by Broido et al.~\cite{broido2019scale}:
evidence of scale-free properties exist either when no alternative heavy-tailed distribution is relatively better than the power law or when the power law is a plausible model for the distribution.
In the former case, the network exhibits \textit{Super-Weak} scale-free structure. 
In the latter, evidence of scale-free properties is said to be \textit{Weak} if the tail of the distribution contains at least 50 nodes, and \textit{Strong} if also $ 2 < \hat \alpha < 3 $ holds. 
We start by estimating the parameters $\hat \theta  = (\hat k_{min}, \hat \alpha)$ by minimizing the Kolmogorov–Smirnov distance between empirical and fitted data for $\hat k_{min}$, and exploit it to estimate $\hat \alpha$ through the method of maximum likelihood estimation~\cite{clauset2009power}.
We then conduct a goodness-of-fit test via a bootstrapping procedure ($N = 5,000$). The resulting p-value indicates if the power law is a plausible fit ($p \geq 0.1$) for the empirical data or not.
Finally, we conduct a log-likelihood ratio ($\mathcal{R}$) test to compare the power law fit against other heavy-tailed distributions (i.e., the Exponential, the Lognormal, and the Weibull). A positive value indicates that the power law distribution is favored over the alternative, and the statistical significance is supported by a p-value that indicates if the hypothesis $\mathcal{R} = 0$ is rejected ($p < 0.1$) or not ($p \geq 0.1$).

% Findings

\begin{table}
    \centering
    \caption{Likelihood ratio and p-value. None of the reported heavy-tailed distributions is favored over the power law.}
    \label{tab:powlaw_comparisons}
    \begin{tabular*}{\columnwidth}{@{\extracolsep{\fill}}lcc}
%\begin{tabular}{@{\extracolsep{\fill}}lcc}
    \toprule
    & DeFi CA Network & DeFi Protocol Network \\
    \midrule
    Exponential & $\mathcal{R}$: 1.322, p-val: 0.186 & $\mathcal{R}$: 4.753, p-val: 0.000 \\
    Lognormal & $\mathcal{R}$: -0.406, p-val: 0.685 &  $\mathcal{R}$: 0.191, p-val: 0.848 \\
    Weibull & $\mathcal{R}$: 1.122, p-val: 0.262 &  $\mathcal{R}$: 2.742, p-val: 0.006 \\
    \bottomrule
\end{tabular*}
\end{table}

Figure~\ref{plot:deg_dist} shows the power law fit for both networks and their estimated $\hat k_{min}$ and $ \hat \alpha$.
Coherently with other studies on the interaction networks from Ethereum blockchain data~\cite{Lee:2020wa}, $\alpha$ lies around 1.7 and 1.8, thus being slightly smaller than the average values usually found for power law distributions. The hypothesis that a power law distribution is a good fit is not plausible for both networks because p-values are 0.020 and 0.035 for the \CA and Protocol networks, respectively.
Table~\ref{tab:powlaw_comparisons} reports the comparisons with other heavy-tailed distributions and shows that the power law is not significantly favored over the Lognormal distribution for both networks, while it is a better fit than the Weibull and the Exponential for the Protocol network.
In summary, according to the classification proposed in Broido et al.~\cite{broido2019scale}, both networks have \emph{Super-Weak} scale-free properties.
Table~\ref{tab:degree} inspects the tails of the distributions and reports the top 15 \CAs sorted by highest degree: most of the \CAs are associated with a few DEX and lending protocols (\emph{1inch}, \emph{UniSwap}, \emph{0x}, \emph{Instadapp}, \emph{Maker}). We can hypothesize that they are part of DeFi compositions, which we will explore further in subsequent sections.

\begin{table*}[h!]
	\centering
	\caption{First $15$ \CAs by highest degree.}
	\label{tab:degree}
		\begin{tabular*}{\textwidth}{@{\extracolsep{\fill}}lrrrrr}
\toprule
                                   Address &  Label & Protocol &  Degree & In degree & Out degree \\
\midrule
0x00000000000049\dots %46c0e9f43f4dee607b0ef1fa1c  
 &  CHI Token %1inch Network: CHI Token
 &     1inch & \num{2713153} &    \num{305627} &    \num{2407526} \\
0x7a250d5630b4cf\dots %539739df2c5dacb4c659f2488d
  & UniswapV2Router02 %Uniswap V2: Router 2
  &   UniSwap &   \num{56007} &    \num{1711} &     \num{54296} \\
0xc02aaa39b223fe\dots %8d0a0e5c4f27ead9083c756cc2
  & EtherToken-v4 %Wrapped Ether
  &   0x &   \num{54469} &  \num{45129} &    \num{9340} \\
0x5c69bee701ef81\dots %4a2b6a3edd4b1652cb9cc5aa6f
  &  UniswapV2Factory %Uniswap V2: Factory Contract
  &   UniSwap &   \num{46408} &    \num{26576} &     \num{19832} \\
0x2971adfa57b20e\dots %5a416ae5a708a8655a9c74f723
  & Mainnet-InstaIndex %InstaDApp: Index
  & Instadapp &   \num{34497} &     \num{18369} &     \num{16128} \\
0x4c8a1beb8a8776\dots %5788946d6b19c6c6355194abeb
  & Mainnet-InstaList %InstaDApp: List
  & Instadapp &   \num{33551} &    \num{16956} &      \num{16595} \\
0x5ef30b99863452\dots %49bc32d8928b7ee64de9435e39
  & CDP\_MANAGER  % Maker: CDP Manager
  &    Maker &   \num{15300} &      \num{8940} &      \num{6360} \\
0x35d1b3f3d7966a\dots %1dfe207aa4514c12a259a0492b
  & MCD\_VAT % Maker: MCD Vat 
  &     Maker &   \num{15214} &     \num{15214} &       \num{0} \\
0xa26e15c895efc0\dots %616177b7c1e7270a4c7d51c997
  & PROXY\_FACTORY  % Maker: DS Proxy Factory 
  &     Maker &   \num{13718} &         \num{1} &    \num{13717} \\
0x0000000000b3f8\dots %79cb30fe243b4dfee438691c04
  & GST2 Token  % contract that tokenizes gas https://gastoken.io/
  &    Unknown &   \num{13447} &     \num{7644} &    \num{5803} \\
0x11111112542d85\dots %b3ef69ae05771c2dccff4faa26
  & contractAddress %1inch V3
   &     1inch &   \num{12371} &    \num{2073} &    \num{10298} \\
0x6b175474e89094\dots %c44da98b954eedeac495271d0f
  & MCD\_DAI  %Dai Stablecoin 
  &     Maker &   \num{12314} &    \num{12314} &        \num{0} \\
0xdef1c0ded9bec7\dots %f1a1670819833240f027b25eff
  & ExchangeProxy-v4 %0x: Exchange Proxy
  &   0x &   \num{11147} &      \num{1138} &      \num{10009} \\
0x939daad09fc4a9\dots %b8f8a9352a485dab2df4f4b3f8
  & mainnet-v1-InstaAccount % InstaDApp: Account 
  & Instadapp &   \num{10876} &     \num{10876} &        \num{0} \\
0xfd3dfb524b2da4\dots %0c8a6d703c62be36b5d8540626
  & N/A % not found in etherscan
  &    Unknown &   \num{10554} &    \num{1547} &      \num{9007} \\
\bottomrule
\end{tabular*}
	\end{table*}

\subsection{Centrality measures}
The results in the previous section highlight the relevant role of DEXs and lending protocols. 
Network centrality measures are another helpful tool to determine which nodes might implement core functionalities.
We consider the In degree centrality, as we are interested in identifying relevant contracts that other protocols may use in DeFi compositions.
To add further insights, we also provide the results for the Katz and PageRank algorithms.
Katz centrality accounts for the importance of a node's neighbors. It is an extension of the eigenvector centrality that addresses issues arising with directed networks~\cite{newman2018networks} by adding a constant initial weight to each node.
PageRank takes into account the Out degree of nodes to control for the drawback of the Katz algorithm that peripheric nodes might get too high values if linked to a very central node. 
The values of each centrality metric are normalized to the range [0,1].

We find that both networks are dominated by a few nodes with relatively high values (for all centrality measures) with respect to the other nodes; the In degree values are almost always higher than the Katz ones, which in turn are often slightly larger than the PageRank centrality values.
Table~\ref{tab:centr} reports the values for the nodes with the highest centrality in the Protocol (left) and the DeFi CA (right) networks. We show only the first three nodes because the others have relatively smaller values in comparison.
In the Protocol network, the most central nodes are two non-labeled \CAs. 
When considering the ranking of the nodes in the highest 10 positions for at least one centrality measure, 10 DeFi protocols appear in the highest positions, and \textit{Uniswap}, in particular, plays an important role. 
Such protocols are thus heavily used by other non-labeled \CAs in our dataset. \textit{Uniswap}, \textit{0x} and \textit{Maker} have higher centrality values with respect to the other protocols. 
The DeFi CA network is dominated by the \textit{1inch} factory contract mentioned in Section~\ref{sec:ground_truth_data} that deploys CHI tokens. 
Two other nodes with relatively high values are the wETH \CA related to 0x 
and another factory contract associated with \textit{Uniswap}. 
Considering again the nodes ranking in the highest 10 positions for at least one centrality measure, 
\CAs associated to \textit{Instadapp} and \textit{Maker} appear repeatedly. Factory deployer contracts play a major role in the DeFi \CA network. Note that, by definition, such contracts have a high Out degree, as their functional role is to deploy other contracts. Interestingly, the In degree centrality results show thus that they also have a relevant role as recipients of calls by other contracts of the network. 
In conclusion, these results are consistent with the findings of Section~\ref{sub:powerlaw} in showing that DEX and lending protocols play a major role and may be involved in compositions.

\begin{table}[t]
	\centering
	\caption{In degree, Katz and PageRank centrality measures the three most central nodes. 
	For the Protocol network (left), the column Address/Protocol reports the address of non-labeled \CAs or the protocol name associated to the node. 
	For the DeFi CA network (right), the column Protocol\_Address reports the protocol associated to the \CA and the \CA itself.
	}
	{\small
\begin{tabular*}{\columnwidth}{@{\extracolsep{\fill}}lccclccc}%\begin{tabular}{>{\centering\arraybackslash}p{0.1\textwidth}>{\centering\arraybackslash}m{0.1\textwidth}>{\centering\arraybackslash}m{0.07\textwidth}>{\centering\arraybackslash}p{0.13\textwidth}}
	\toprule   
	\multicolumn{4}{c}{Protocol network} &  \multicolumn{4}{c}{DeFi CA network}  \\
	\cmidrule{1-4} 
	\cmidrule{5-8}
	Address/Protocol & In degree  & Katz  & PageRank  & Protocol\_Address  & In degree  & Katz  & PageRank \\
	\cmidrule{1-4} 
	\cmidrule{5-8}
	0x0000000000b3f8\dots & 1  & 1  & 1  & 1inch\_0x00000\dots  & 1  & 1  & 1 \\
	0xcc88a9d330da11\dots & 0.371  & 0.168  & 0.111  &  0x\_0xc02aa\dots & 0.148  &  0.107 & 0.053 \\
	UniSwap & 0.313  & 0.176  & 0.092  & UniSwap\_0x5c69b\dots &  0.087 &  0.064  &  0.036 \\
	\bottomrule
\end{tabular*}
}
	\label{tab:centr}
\end{table}

\subsection{Reciprocity and assortativity}
Next, we look at two measures that provide information on the relationship between nodes and their neighbors, that is, reciprocity and assortativity.
Reciprocity is the likelihood that nodes are mutually linked.
Values range from 0 to 1, the former meaning that the network is purely unidirectional, the latter indicating that all links are reciprocated.
For both the DeFi CA and the Protocol networks, the values (respectively 0.234 and 0.215) are similar to the one reported in~\cite{Lee:2020wa}, and we follow their interpretation that the presence of reciprocated links is a potential sign of composability, as it shows that smart contracts tend to rely often on each other.
The lower value obtained for the Protocol network could be explained by the presence of many non-labeled (non-protocol-specific) \CAs.
If we further reduce the Protocol network by removing all non-labeled \CAs, obtaining a graph abstraction of 23 nodes, the reciprocity (0.677) is much higher, indicating that protocols interact with each other more often and in a bidirectional way, a sign that compositions exist. 
Assortativity is a metric that indicates whether nodes with similar degrees tend to interact with each other ($1 > \rho > 0 $), or if nodes with high degrees interact more with low degree nodes ($ 0 > \rho > -1 $). Consistently with previous results on the Ethereum transaction network, both networks are disassortative (-0.473 for the DeFi CA network and -0.262 for the Protocol network), indicating heterogeneity and a sign that CAs with high degree are leveraged by many other CAs with a less relevant role in the ecosystem. As shown above, such nodes are often associated with DEX and lending protocols. 

\subsection{Components}

% Motivation

Reciprocity shows that protocols interact bidirectionally with accounts related to other protocols.
We thus look at metrics providing further insights on how the (code accounts of) different protocols fall into distinct disconnected components.
We distinguish between \textit{weakly} connected components, in which all the nodes are connected by a path independently of the directions of the edges, and \textit{strongly} connected, which considers the edge direction.

% Findings

\begin{figure}
     \centering
     \includegraphics[width=0.75\columnwidth]{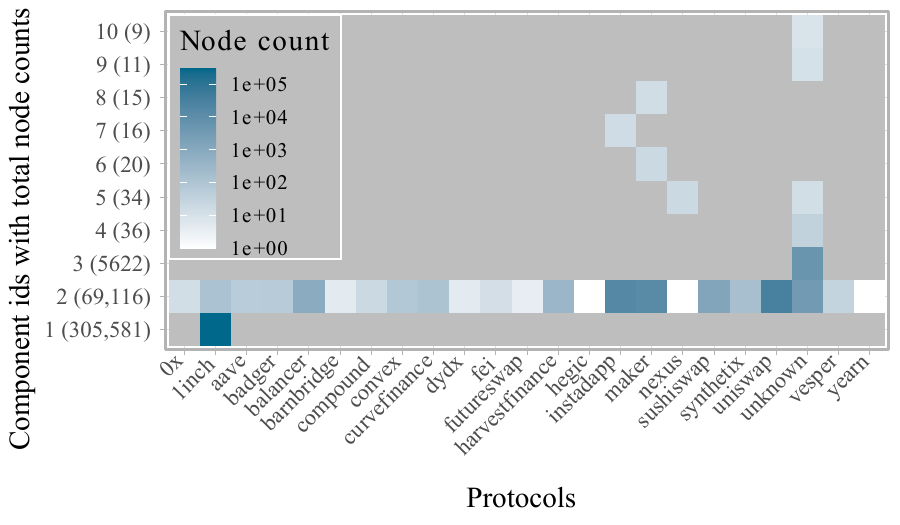}
     \caption{Heatmap showing how the addresses associated to different protocols fall into the ten largest strongly connected components. The largest component is uniquely composed of 305,581 \emph{1inch} addresses, while the second collects the vast majority of protocols. Smaller components identify addresses of protocols that do not interact outside of the protocol itself.}
     \Description{Heatmap showing how the addresses associated to different protocols fall into the ten largest strongly connected components. The largest component is uniquely composed of 305,581 \emph{1inch} addresses, while the second collects the vast majority of protocols. Smaller components identify addresses of protocols that do not interact outside of the protocol itself.}
     \label{fig:ca_heatmap}
\end{figure}

For the Protocol network, we find that the largest weakly connected component is equal to the entire network, while for the \CA network, only 34 nodes are outside of the largest component. The remaining nodes fall into 16 components, with a few nodes each.
Table~\ref{tab:comps_summary} lists the three largest strongly connected components. 
By comparing the number of edges and nodes, we notice that the second-largest component of both the Protocol and the \CA network is denser than the other larger components. Additionally, in Figure~\ref{fig:ca_heatmap} we illustrate how the \CAs belonging to different protocols map to the ten largest strongly connected components of the \CA network. Interestingly, the second-largest component also encompasses the vast majority of protocol interactions. While the largest component is entirely composed of \CAs associated with the \emph{1inch} protocol, in the second-largest component, we find addresses of all the analyzed protocols except for \emph{RenVM}, which is not present in any of the reported large components. We also find that all the protocols fall into the second-largest strongly connected component regarding the Protocol network.
This analysis shows that interactions among protocols primarily occur in a single, large component that is more interconnected than average. Notably, such interactions might indicate the existence of compositions due to the overlapping transaction structure of multiple protocols.

\begin{table}
    \centering
    \caption{Description of the three largest strongly connected components. For both networks the pattern is fragmented, but interestingly the second largest strongly connected components are remarkably more interconnected, indicating that nodes in these components interact with many other nodes, a prerequisite for composition.}
    \label{tab:comps_summary}
    {\footnotesize
    \begin{tabular*}{\columnwidth}{@{\extracolsep{\fill}}lrrrrrrr}
\toprule
        &     & \multicolumn{2}{c}{Largest} & \multicolumn{2}{c}{2{nd} largest}  & \multicolumn{2}{c}{3{rd} largest}  \\
 %\cmidrule{3-4} \cmidrule{6-7}  \cmidrule{8-9} \cmidrule{10-11}   
   & \# Comp. & Nodes & Edges  & Nodes  & Edges  & Nodes  & Edges  \\
\midrule
Contract &      \num{2155707} &        \num{305581} &        \num{611160} &         \num{69116} &        \num{370833} &          \num{5622} &         \num{11242} \\
 Protocol &            \num{33832} &          \num{5622} &         \num{11242} &          \num{3948} &         \num{14264} &            36 &            71 \\
\bottomrule
\end{tabular*}}
\end{table}

\subsection{Community Detection}

% Motivation
One could naively assume that \CAs associated with specific DeFi protocols form communities in the Code Account network. However, the previous results suggest that the network topology reflects DeFi compositions at the level of the community structure. We thus measure how effectively different community detection algorithms detect protocols in the DeFi \CA network.

% Method

We follow the approach of Yang et al.~\cite{yang2016comparative}, who provide guidelines for selecting community detection algorithms depending on the size of the network. We analyze the weakest largest component in its unweighted and undirected version with non-overlapping communities using four different algorithms: multilevel or Louvain~\cite{blondel2008fast}, label propagation~\cite{raghavan2007near}, leading eigenvector~\cite{newman2006finding}, and Leiden~\cite{traag2019louvain}.
Using the labeled addresses in our \textit{ground truth} dataset, we can verify to what extent $\hat{C}$, the set of communities identified by partitioning algorithms, correspond to $P^{\ast}$, the set of ground truth communities defined by the individual protocols.
We quantify their performance through the \textit{normalized mutual information (NMI)}, a benchmark measure in the literature \cite{danon2005comparing,lancichinetti2009benchmarks} that quantifies the similarity between the ground truth communities and the identified communities. In addition, we provide two additional measures: the ratio $\hat{C} / P^{\ast}$ for the accuracy of the number of identified communities and the F1 score.
We compute the latter similarly to \cite{yang2012community}: first, for each protocol $P_{i} \in P^{\ast}$ we identify the detected community $ C_{j} \in \hat{C}$ that maximizes the F1 score. Then, we report average precision, recall, and F1 scores over all communities $P_{i} \in P^{\ast}$. Note that we compute the above metrics only on the labeled \CAs.
%
% Findings
%
\begin{table}[]
    \centering
    \caption{Performance metrics for the community detection algorithms. Low F1 Scores indicate either that the algorithms poorly identify communities, or that the network topology reflects a more complex organization at the mesoscopic level.}
    \label{tab:communities_summary}
    {\footnotesize
    \begin{tabular*}{\columnwidth}{@{\extracolsep{\fill}}lcrrrrrrr}
\toprule
          Algorithms & Communities  & Precision &  Recall & F1 Score &     NMI & $\hat{C} / P^{\ast}$ \\
\midrule
    Louvain &       14  &    0.3896 &  0.7181  &   0.2917 &  0.9241 &             0.6087 \\
     Leiden &       10 &    0.3021 &  0.8589 &        0.2879 &   0.9620 &             0.4348 \\
 Label prop. &      53 &     0.7107 &  0.6009 &       0.4892 &  0.9404 &              2.3043 \\
Eigenvector &         4 &    0.1696 &   0.9070 &       0.1776 &  0.9495 &             0.1739 \\
\bottomrule
\end{tabular*}}
\end{table}
The second column of Table~\ref{tab:communities_summary} reports the total number of communities that include labeled \CAs.
The NMI is high for all the protocols, indicating that overall the algorithms correctly partition the network: indeed, all algorithms cluster together the \CAs created by the \emph{1inch} \textit{deployer} contract, and \emph{1inch} is by far the largest ground truth community in terms of labeled accounts.
On the other hand, the low F1 scores (0.18-0.49) result from a small set of misclassified ground truth communities (e.g., \emph{Compound}, \emph{DyDx}, \emph{Fei}). Upon closer inspection, we noticed that some protocols map entirely into a few communities dominated by larger protocols (such as \emph{UniSwap} or \emph{Maker}), negatively impacting precision, while others are split into different communities, affecting recall. \emph{1inch} itself has a non-marginal number of addresses that map into other communities.

In summary, we see that algorithms work well, with NMI scores above 0.92. However, when considering the imbalance in our dataset (precision, recall), we find that known community detection algorithms cannot effectively identify protocols as distinct communities, but rather indicate protocol composition patterns. 
The identified community structure reflects a different organization in which protocols are entangled.

% !TeX root = ../main.tex

\section{Measuring DeFi Compositions}\label{sec:composition}

After analyzing the macroscopic network perspective, we now address the microscopic trace level, where we identify and extract building blocks,
i.e. recurring patterns of internal traces induced by protocol-specific \CAs that are found as subpatterns within different transactions. The  building block detection can help better understand DeFi compositions and identify a variety of risks. We consider a detailed risk analysis to be future work, but can motivate some sources of risk: for example, if security vulnerabilities are identified in underlying building blocks, they can propagate to higher levels and pose a risk to other DeFi protocols. Atzei et al.  \cite{Atzei2017SurveyAttacks} analyze the security vulnerabilities of Ethereum code accounts and attacks that exploit them. Legal issues may arise, including licensing issues, thereby limiting usability in other protocols. This phenomenon	also exists in traditional software\footnote{\url{https://www.techradar.com/news/this-popular-code-library-is-causing-problems-for-hundreds-of-thousands-of-devs}}. Finally, the technical evolution of a blockchain can also have an impact on the efficiency or security of an existing building block, and here too it is important to identify which protocols are affected.

Thus, we propose an algorithm to extract the possibly nested structure of DeFi protocol calls, which may also be used by other DeFi protocols. In contrast to recent works, that have discovered and exposed DeFi compositions, we provide a systematic, automated mechanism to explore them by using  building block extraction. 
We then assess the most frequent building blocks our algorithm identifies and illustrate possible DeFi compositions and show how the DEX aggregator \emph{1inch} and the \emph{Instadapp} protocols use multiple such building blocks of other protocols. Further, we flatten the nested structure of building blocks and study the interaction of DEX and lending services. Finally, we present in a case study the dependencies of DeFi protocol on stablecoins, by using our extracted building block.

\begin{figure}[b]
	\centering
	\begin{subfigure}[b]{0.32\textwidth}
	\includegraphics[width=1\linewidth]{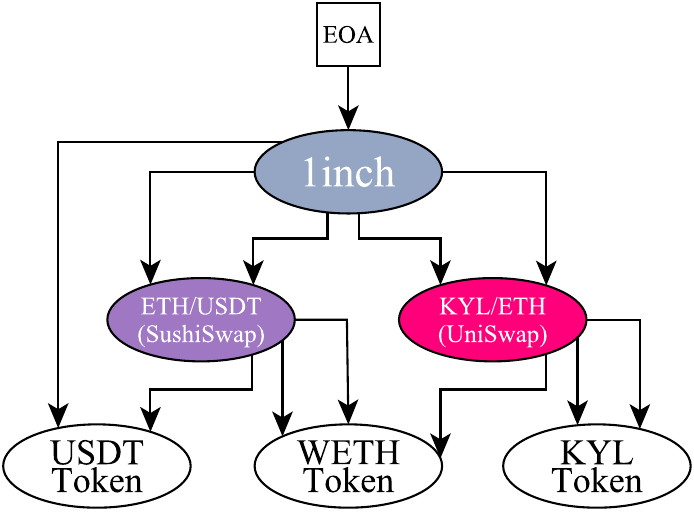} %height=6.4cm
	\caption{Original transaction trace graph of the composition as shown in Figure~\ref{fig:composition_example}.}
	\label{fig:buildingblock-p1}
	\end{subfigure}\hfill
	\begin{subfigure}[b]{0.32\textwidth}
	\includegraphics[width=1\linewidth]{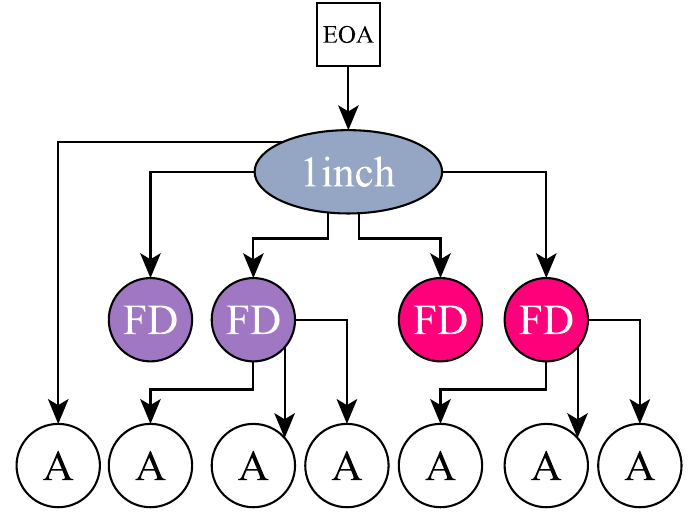} %height=6.4cm
	\caption{Conversion to execution tree, renaming factory deployed contracts and assets.}
	\label{fig:buildingblock-p2}
	\end{subfigure}\hfill
	\begin{subfigure}[b]{0.32\textwidth}
	\includegraphics[width=1\linewidth]{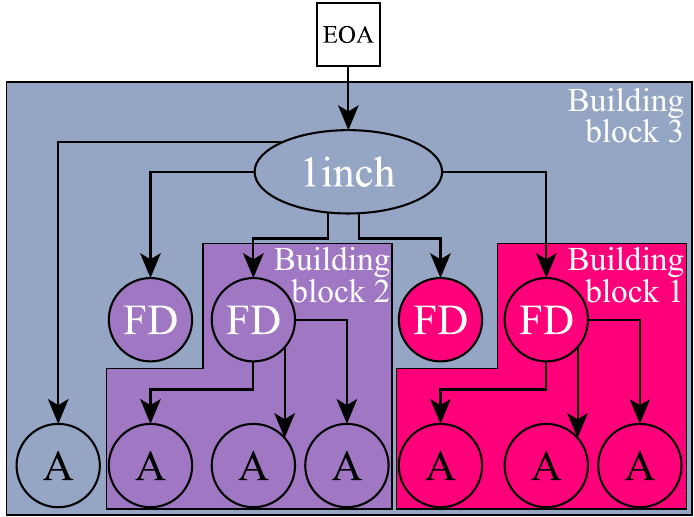}
	\caption{Identification of general building blocks from protocol nodes with subtraces.}
	\label{fig:buildingblock-p3}
	\end{subfigure}
	\caption{A high-level illustration of the building block extraction algorithm. Subfigure~\ref{fig:buildingblock-p1} represents the input composition. This graph is then converted into an execution tree as shown in Subfigure~\ref{fig:buildingblock-p2}, such that each node can only have one incoming edge, requiring the duplication of nodes. In addition, the underlying assets (tokens) and factory deployed contracts are renamed. In this example, the trading pair contracts are factory deployed (FD). This allows for the identification of generalized building blocks, as each trading pair only differentiates itself by the specific assets it is dealing with. The result of the building block extraction is then shown in Subfigure~\ref{fig:buildingblock-p3}, and is the result of a bottom-up processing of the tree, selecting subtrees of known protocol nodes. See Algorithm~\ref{alg:buildingBlocks} for more details.}
	\label{fig:buildingblock-extraction}
\end{figure}

% !TeX root = ../main.tex

\subsection{Building Block Extraction Algorithm} \label{sec:buildingBlocks}
In order to detect building blocks, we treat individual transactions as trees of execution traces, that is, as
an abstraction where the external and all the internal transactions are represented as an edge to a new node (thus, the same \CA appears multiple times if executed more than once). We break the trees into subtrees, starting from the tree's leaves, and identify a building block whenever we encounter a node that is part of a protocol. If multiple protocol nodes exist in a tree, the building blocks can be composed of one another. To obtain the nested structure, we create a hash of each building block and use those hashes to chain nested tree structures. Figure~\ref{fig:buildingblock-extraction} illustrates the process from a high-level perspective. Subfigure~\ref{fig:buildingblock-p1} represents the input, which corresponds to the original transaction trace graph which we've also shown in the introductory Figure~\ref{fig:composition_example}.
We aim to identify building blocks that execute the same logic despite being different instances involving different addresses (i.e., a swap with different tokens). We preprocess and generalize the execution trace trees as follows:

\textbf{Preprocessing:} In contrast to a graph, like in Figure~\ref{fig:buildingblock-p1}, an execution tree can have the same node appearing multiple times as a leaf node, effectively having no cycles. Each edge has a trace ID, determining the order of the calls. If a contract address appears in a trace that has been deployed by a factory, we rename it to \texttt{\$protocol-DEPLOYED}. Furthermore, we rename all contract addresses as \texttt{ASSET}, which fulfill the criteria that their smart contract code contains the standard ERC20 token method signatures, and if within the trace, the token contract is called with one such method. The result of these preprocessing steps is shown in Figure~\ref{fig:buildingblock-p2}. This preprocessing assumes that factory deployed contracts and ERC20 token contracts provide similar functionality. This allows us to generalize the traces, as many similar interactions with various standardized tokens become identical.

\begin{algorithm}
	\setstretch{1.25} % increase line spacing
    \SetKwInOut{KwInputs}{Inputs}
    \SetKwInOut{KwOutputs}{Outputs}
    \SetAlgoLined
    \KwInputs{(1) Directed, attributed transaction trace tree $G(V,E,t,m)$ with functions $t:E \rightarrow \mathbb{N}$ assigning a unique trace ID, and $m:E \rightarrow \mathbb{N}$ assigning a method ID on the edges of the tree, \\(2) protocol vertices $V_P$}
    \KwOutputs{Lists of building blocks $L_B$, and hashes $L_B^h$}
    $L_B \leftarrow (\ )$\tcp*{Init. list of building blocks}\label{alg:init_bb}
    $L_B^h \leftarrow (\ )$\tcp*{Init. list of building block hashes}\label{alg:init_bb_hash}
    $L_e \leftarrow (uv) | \forall uv \in E: v \in V_P$\tcp*{Edges to protocol nodes}\label{alg:protocol_edges}
    \tcp{For each edge to a protocol, get subtree}
    $L_E^p \leftarrow (E_i) | $ edges reachable from $e_i$ for each $e_i \in L_e$\;\label{alg:subsequent_edges}
    $L_G^p \leftarrow (G[E_{i}])\ |\ \forall E_{i} \in L_E^p$\tcp*{edge induced subtrees}\label{alg:get_subtrees}
    $L_G^p \leftarrow$ filter($L_G^p$, by=tree-depth, minimum=2)\;\label{alg:filter_trees}
    $L_G^p \leftarrow$ sort($L_G^p$, by=tree-depth, how=ascending)\;\label{alg:sort_trees_by_depth}
    \For(\tcp*[f]{for each subtree}){$G_S(V_S, E_S, t, m) \in L_G^p$}{\label{alg:for_each_subtree}
    \tikzmk{A}\tcp{Compute building block hash with $V_{s}', D_{s}, M_{s}$}\label{alg:hash_comment_line}
    $E_{S}' \leftarrow$ sort($E_S$, by=$t(E_S)$, how=ascending)\tcp*{Sort edges}\label{alg:sort_edges}
    $V_S' \leftarrow (v_1,...,v_n) = v | \forall uv \in E_S': v\in V_S$\tcp*{Vert. list}\label{hash_vertex_part}
    $D_{S} \leftarrow deg_{out}(v)|\forall v \in V_S'$\tcp*{Outdegree list}\label{hash_outdeg_part}
    $M_{S} \leftarrow m(E_{S}')$\tcp*{Method ID list}\label{hash_method_part}
    $h_S \leftarrow$ sha256hash(stringify$(V_{s}', D_{s}, M_{S})$)\;\label{hash_all_parts}  \tikzmk{B}\boxit{gray}
    $B_S \leftarrow G[V_S']$\tcp*{B. block as vertex induced subtree}\label{alg:get_BB}
    replace(what=$G_S$, in=$G$, with=$h_S$)\;\label{alg:replace}
    $L_B \leftarrow L_B \mathbin\Vert B_S$\tcp*{Append building block}\label{alg:append_BB}
    $L_B^h \leftarrow L_B^h \mathbin\Vert h_S$\tcp*{Append building block hash}\label{alg:append_BB_hash}
    }
    \KwRet{$L_B, L_B^h$}\label{alg:return}
\caption{Building Block Extraction}
\label{alg:buildingBlocks}
\end{algorithm}

\textbf{Algorithm~\ref{alg:buildingBlocks}} takes as input a transaction trace tree $G(E,V,t,m)$ with two edge attributes: the trace ID $t$, indicating the order of execution, and $m$, indicating the method ID of the executed call.
The second input is a list of seed protocol nodes, such as those described in Section~\ref{sec:ground_truth_data}. The algorithm outputs a list of building blocks and hashes of such building blocks.
We first setup the output variables in lines~\ref{alg:init_bb}--\ref{alg:init_bb_hash}.
We then find edges to the protocol nodes in line~\ref{alg:protocol_edges} and extract all further reachable edges of these to obtain edge-induced subtrees in lines~\ref{alg:subsequent_edges}--\ref{alg:get_subtrees}.
We filter them in line~\ref{alg:filter_trees} to include only those with a minimum depth of 2, such that the protocol node has to make further calls.
In line~\ref{alg:sort_trees_by_depth}, we sort the list of subtrees ascendingly based on their depth. This means small trees are at the beginning of the list, and large trees that may contain these smaller trees are at the end.
For each subtree (line~\ref{alg:for_each_subtree}), we compute a hash in lines~\ref{alg:hash_comment_line}--\ref{hash_all_parts}, highlighted in gray, akin to a tree kernel.
To compute the hash, we first sort the subtree's edges by order of execution in line~\ref{alg:sort_edges}, and then extract the target vertices of each edge in line~\ref{hash_vertex_part}, essentially excluding the original calling node, which could be different in each transaction. For each of those vertices, we compute the outdegree (line~\ref{hash_outdeg_part}), and also determine the method ID for each edge (line~\ref{hash_method_part}).
The hash is then computed from the three aforementioned properties in line~\ref{hash_all_parts}.
Using the target vertices, we retrieve the building block from the original tree (line~\ref{alg:get_BB}), which may contain leaf nodes of building block hashes as replacing subtrees in line~\ref{alg:replace} can lead to nested building blocks. Finally, we append building block and hash to their lists in lines~\ref{alg:append_BB}--\ref{alg:append_BB_hash}. Once all subtrees are processed, the lists are returned in line~\ref{alg:return}.

An example of the algorithm's result can be seen in Figure~\ref{fig:buildingblock-p3}, showing three building blocks, one each from \textit{SushiSwap}, \textit{UniSwap} and \textit{1inch}. Note that the building block of 1inch contains the other two building blocks.

\begin{figure}
	\centering
	\includegraphics[width=\linewidth]{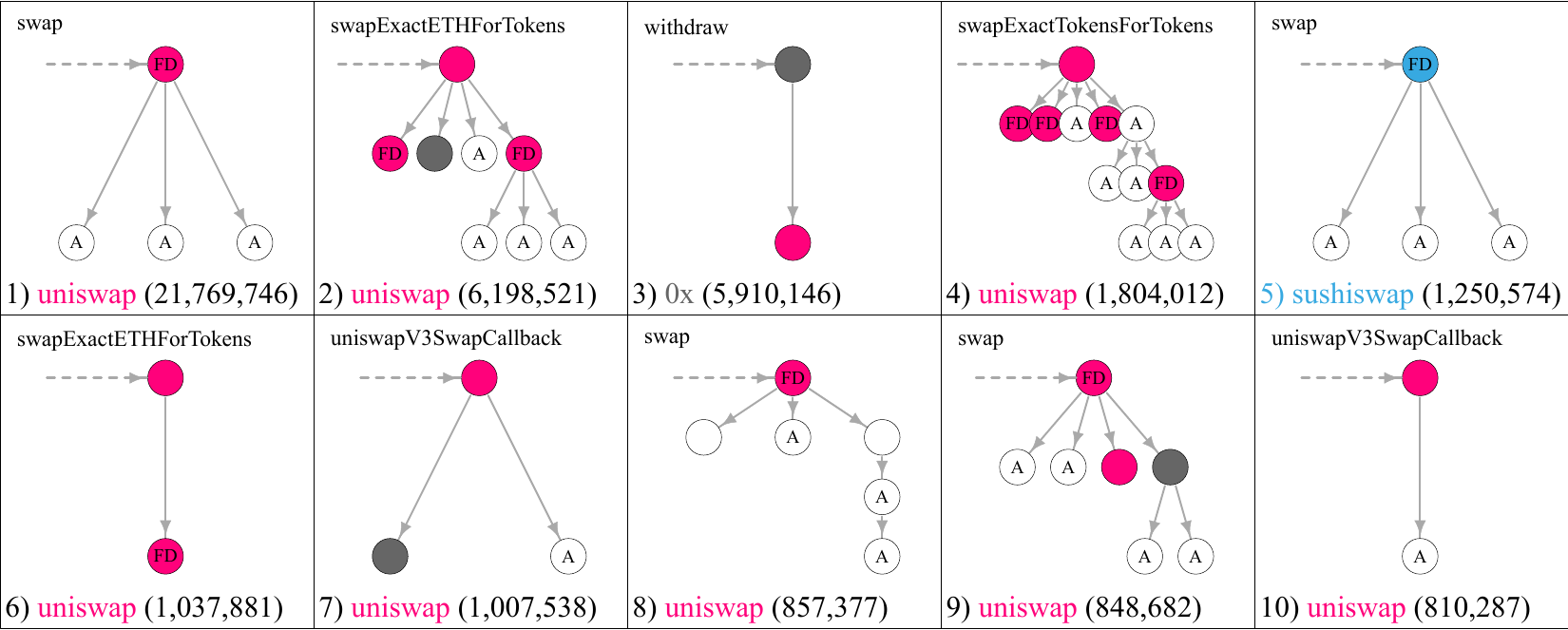}
	\caption{The 10 most frequently observed building blocks by called root method, root protocol and count. Nodes marked with \texttt{FD} are generalized factory deployed contracts and those marked with \texttt{A} are ERC20 assets. The majority of these building blocks originate from \emph{UniSwap}. Note that block 1 of \emph{UniSwap} is equivalent to number 5 of \emph{SushiSwap}. This makes sense, as \emph{SushiSwap} is a fork of \emph{UniSwap}. Number 1 is contained in building blocks 2, and 4 -- illustrating an internal composition within the same protocol. Building block 3 represents the withdrawal of Wrapped Ether (\emph{WETH}) and is associated to the protocol \emph{0x}.
	Also note that several root methods are identical, yet can lead to different types of building blocks.
	}
	\label{fig:buildingblocks}
\end{figure}

% !TeX root = ../main.tex

\subsection{Building Block Analysis}\label{sec:block_analysis}

We execute the algorithm on all transactions in our dataset, together with the set of DeFi protocols in our labeled extended seed set (cf. Section~\ref{sec:data_methods}). We can then count the retrieved building blocks by their hashes, understand their composition, and visualize them.
Figure~\ref{fig:buildingblocks} illustrates the top 10 most frequently observed building blocks, of which eight belong to \emph{UniSwap}. The most frequent building block is a \emph{UniSwap} swap, with more than 21 million occurrences. As \emph{UniSwap} is one of the most popular DeFi protocols, and token swaps are its main functionality, this result shows that the building block extraction is meaningful. We further observe that the swap building block is reoccurring and contained in other patterns that appear frequently.
Another relevant block is related to \emph{0x}'s Wrapped Ether (\emph{WETH}), which in our context is not classified as an asset due to its' use of withdrawal, a non ERC20 function.
In the following, we will provide more insights into the nested structure from different perspectives and discuss their interpretations.

\subsubsection{Protocol Building Block Composition}\label{sec:521}
\begin{figure}
\centering
\includegraphics[width=1\linewidth]{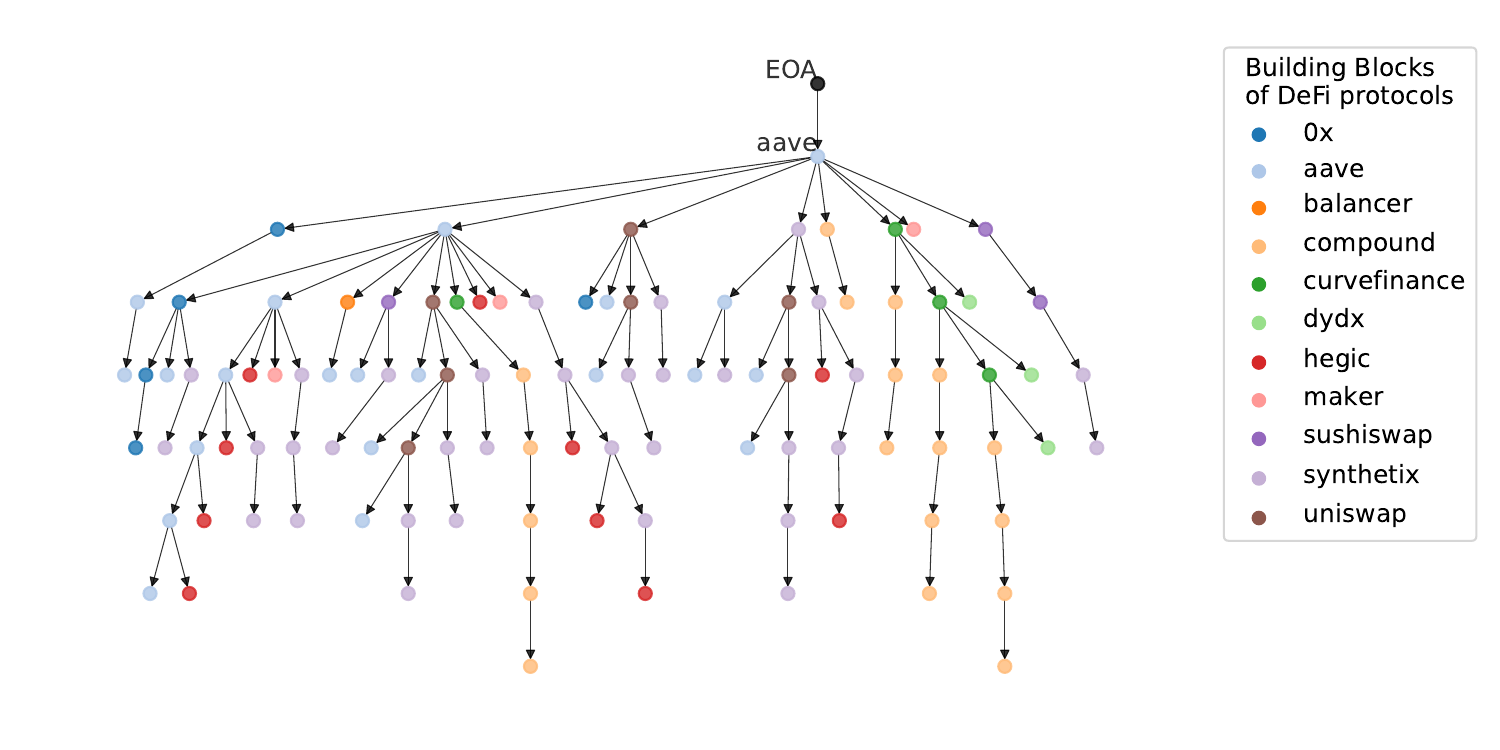}
\caption{Illustrating the composition space of \emph{Aave} as a network tree. Each node represents a building block, each link a possible nested building block, extracted from all transactions to \emph{Aave}. We observe for this protocol a maximum depth of seven nested DeFi building block levels.}
\label{fig:bb_tree_aave}
\end{figure}

Starting from the execution tree structure of each trace, the algorithm identifies  subtrees. Those building
blocks obtained from Algorithm~\ref{alg:buildingBlocks} can contain leaves with hashes that point to other building blocks, leading to a nested structure that still preserves the primary tree structure of the traces. But a single transaction only represents a small snapshot of the entire tree of possible compositions.
For a comprehensive image of the DeFi protocols composition space, we have to consider multiple transactions. To observe the space of all possible compositions, we construct a network of overlapping building block trees for all transaction of the same initial (external) DeFi protocol. For an illustrative example, we used the extracted building block structures of all transaction to \emph{Aave}. The network still conserves the tree structure, where each node represents a building block and each link a nested composition, observed in the transactions. Figure \ref{fig:bb_tree_aave} shows the \emph{Aave} network and illustrates its multiple nested levels. Starting from the top with external transactions from EOAs to \emph{Aave}, a variety of paths and compositions can be seen, presenting the space of all possible compositions, observed from existing transaction data. Nevertheless, this network illustration doesn't provide a comprehensive picture of the volume (i.e. number of appearances) of those compositions and the number of branches, when a building block calls multiple sub-blocks.

\begin{figure}
	\centering
	\begin{subfigure}[b]{0.49\textwidth}
	\includegraphics[width=1\linewidth]{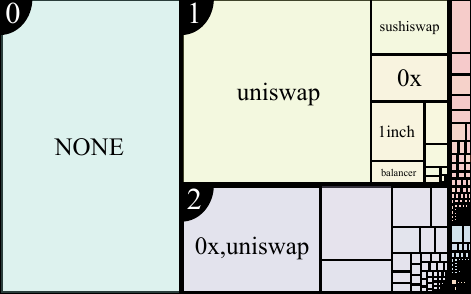} %height=6.4cm
	\caption{1inch}
	\label{fig:treemap_1inch}
	\end{subfigure}\hfill
	\begin{subfigure}[b]{0.49\textwidth}
	\includegraphics[width=1\linewidth]{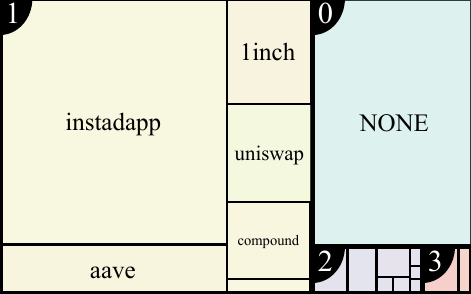}
	\caption{instadapp$\rightarrow$instadapp$\rightarrow$instadapp$\rightarrow$instadapp}
	\label{fig:treeMap_instadapp_depth4}
	\end{subfigure}
	\caption{Inspecting the potentially nested building blocks used by the first level of \emph{1inch} (left) and the fourth level of \emph{Instadapp} (right). The size of each box represents the share of building blocks assigned to one or more unique protocols. For \emph{1inch} transactions, at the first nested level, about a third of the used building blocks are of one (chiefly other) protocol (yellow boxes). An even bigger fraction can be observed for \emph{Instadapp} but in the fourth nested building block level.}
	\label{plot:treemap}
\end{figure}

We can inspect for each building block the set of contained protocols and the volume of their appearances:
 the treemaps in Figure~\ref{plot:treemap} illustrate the shares of protocols appearing in the building block structure of a specific nested level.

In Figure~\ref{fig:treemap_1inch} we observe the volume of building block calls and associated protocols in the first level for the protocols \emph{1inch}.
The largest fraction are external transactions that do not contain any other building blocks; this is captured by the box labeled as \emph{NONE}. All other boxes show instead the share of transactions in which one or multiple DeFi service building blocks are nested.
We group them using different colors based on the number of unique, distinct protocols that are called in the subsequent building blocks of this level. For instance, yellow boxes indicate the fraction of transactions in which the appearing nested building blocks in the first level are associated to one single DeFi protocol, while blue boxes represent the fraction in which the building blocks in the first level are associated to two different protocols. We further observe portions of transactions that contain building blocks assigned to more than two protocols within the first nested level. 

Moreover, the treemap in Figure~\ref{fig:treeMap_instadapp_depth4} show branches in a deeper level within \emph{Instadapp} transactions. In the fourth level of self-compositions, besides the fraction that does not contain any further block (\emph{NONE}), an even bigger share of building blocks appear that are associated to one single DeFi protocol. We also inspect again the existence of building blocks associated to two and more protocols.

These two illustrations in Figure~\ref{plot:treemap} give insights to our systematical investigation on compositions, and show that looking only to selected compositions or single nested levels of DeFi compositions would return a partial picture: interactions among protocols can be iteratively nested one within each other and can take place in deeply nested levels. Therefore a further investigation to disentangle and flatten the nested structure is needed.

\subsubsection{Flattening Composition Hierarchies}

We then want to investigate to what extent the DeFi protocols leverage other protocols to provide their services. That means, we want to identify a mapping of top-level protocols to any of the building blocks they make use of, whether deeply nested or not. To get an overall picture of the DeFi compositions, we flatten the nested building block structures.

\begin{figure}
	%Do not try to scale figure in .tex or you loose font size consistency
	\centering
	\includegraphics[width=1\linewidth]{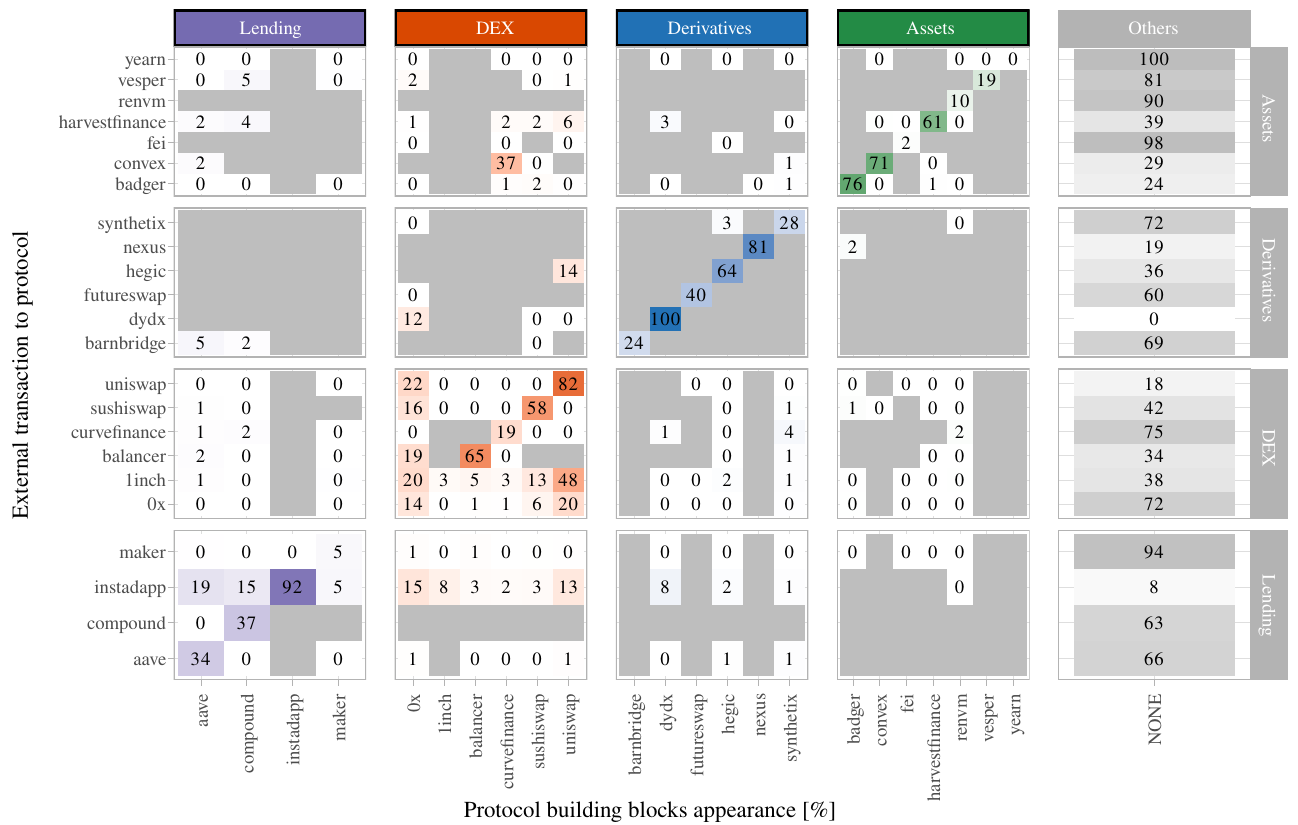}
	\caption{Appearances of DeFi service building blocks across protocols. The numbers indicate the percentage of transactions in which a building block of a certain protocol is contained. The use of multiple DeFi services can be observed for DeFi aggregation protocols, like \emph{Instadapp}, \emph{1inch} and \emph{0x}.}
	\label{plot:fig_pro}
\end{figure}

In each transaction, we follow the cascade of nested building blocks and create a mapping from the contained protocol of building blocks to the original DeFi protocol that the external transaction was sent to (the root protocol). If mappings appear multiple times over different transactions, we aggregate them. For each root protocol, we can then compute the frequency of associated protocols to contain building blocks over all transactions. The result is a measure that indicates, for a given root protocol, the probability that a certain building block of a DeFi protocol appears anywhere in the (nested) building block structure.
In Figure~\ref{plot:fig_pro} we show the building block appearances of lending, DEX, derivatives and asset protocols with a heat map.
Each row corresponds to the external calls to a specific protocol, and the row entries indicate the frequencies of the occurrence of a protocol's building blocks.
The relative share measurement is the fraction of internal building blocks based on the number of external transactions. We notice that the \emph{NONE} category indicates the share of transactions for which no building blocks have been found.
Most protocol interactions exist within each protocol, visible by the highlighted diagonal elements. This pattern is especially remarkable for derivative protocols. Consider, e.g., \emph{dYdX}: all external transactions directed to it contain at least one \emph{dYdX} building block.
However, DeFi aggregation protocols such as \emph{Instadapp}, \emph{1inch}, and \emph{0x} in particular show extensive use of other DeFi services and thus frequent occurrences of DeFi compositions. This indicates Algorithm~\ref{alg:buildingBlocks} works as intended, as, by definition, aggregation protocols must call other protocols. 
The frequent appearance of the \emph{0x} protocol can be attributed to the popular \emph{Wrapped Ether} token and its \emph{withdraw} pattern, already observed and shown in Figure~\ref{fig:buildingblocks}.
Further, we note that second to \emph{0x}, \emph{UniSwap} building blocks appear in most transactions to the protocols shown in Figure~\ref{plot:fig_pro}.
Derivatives protocols have instead little or no further interactions with other protocols, as shown in the row associated with derivatives in the matrix of heat maps, as well as the assets protocols that do not interact heavily with other protocols.

% !TeX root = ../main.tex

\subsection{Case Study: A hypothetical run on the Tether}\label{sec:case_study}

In May 2022, we witnessed the collapse of the Terra ecosystem and its stablecoin TerraUSD (UST), which maintained its peg to the US Dollar through an arbitrage mechanism with the token LUNA. This triggered a so-called stablecoin-run and destroyed over 30B USD of value within a single week.
Motivated by this recent demonstration of systemic risk associated with stablecoins, we apply our building block extraction and analysis methods to measure how a hypothetical run on the stablecoin Tether (USDT)\footnote{\textit{0xdAC17F958D2ee523a2206206994597C13D831ec7}}, which is the most widely adopted stablecoin in Ethereum, would affect known DeFi protocols based on building block dependencies. We distinguish between \emph{direct} dependencies, where USDT is an explicit part of a building block, and \emph{indirect} dependencies, where USDT appears somewhere in its' nested building blocks.
Starting with the most frequent building blocks (see Figure~\ref{fig:buildingblocks}), we analyzed the occurrence of USDT in the regularly used sub-patterns of transactions. We detected USDT in \num{10.6}\% of `swap' building blocks from \emph{UniSwap} (1) and \num{16.2}\% from \emph{SushiSwap} (5). For the `swapExactTokensForTokens' building block from \emph{UniSwap} (2), we find an even higher direct occurrence of \num{22.7}\% and an indirect dependency of further \num{21.2}\% with the nested block structure, containing the before mentioned `swap' building blocks from \emph{UniSwap} (1).

\begin{figure}
	\centering
	\includegraphics[width=1\linewidth]{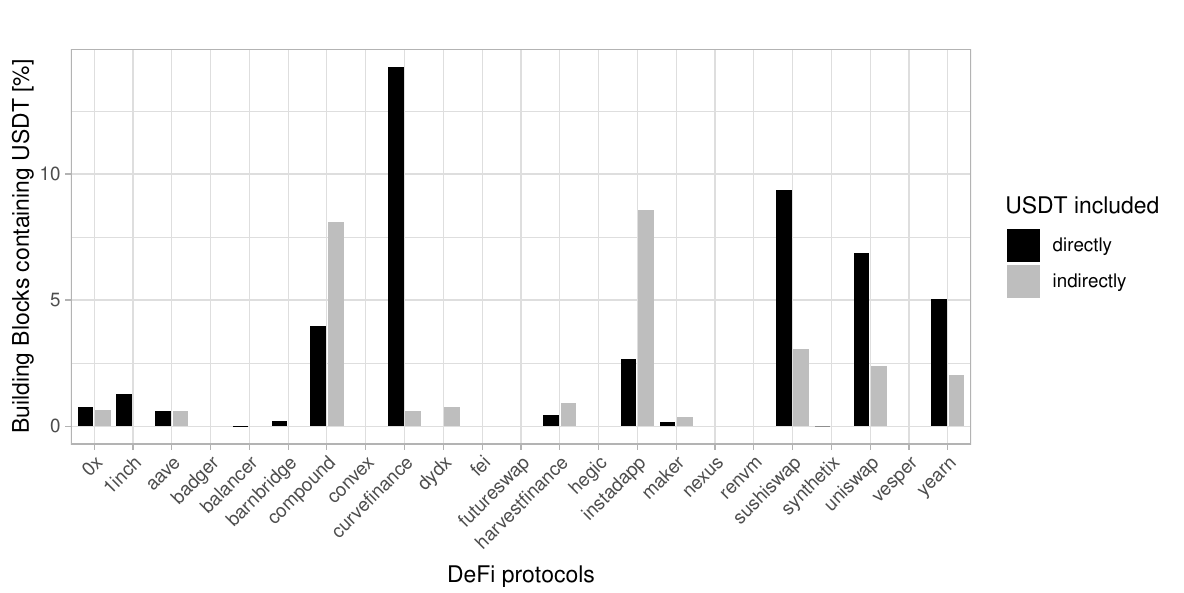}
	\caption{Dependencies of building blocks on the USDT crypto asset for each DeFi protocol. Distinguished between direct included asset or indirect through other nested contained blocks.}
	\Description{ADD DESCRIPTION}
	\label{fig:dep_pro_usdt}
\end{figure}

In order to obtain a broader picture of the dependencies in the DeFi ecosystem, we also analyzed, for each protocol, the fraction of building blocks containing the USDT asset directly or indirectly in more deeply nested blocks. Our results, which are summarized in Figure~\ref{fig:dep_pro_usdt}, show that most protocols have rather low dependencies ($< 10\%$). However, \num{14.2}\% of \emph{Curvefinance} building blocks include the USDT asset directly and also the two DEX protocols \emph{UniSwap} and \emph{SushiSwap} strongly depend on that asset. This is in line with our previous finding that `swap' is the most frequent building block. We further find that \emph{Compound} and \emph{Instadapp} building blocks have in comparison high indirect dependencies on the USDT asset.
These dependencies indicate how a shock in the DeFi ecosystem, such as a run on a stablecoin, could affect DeFi protocols, directly and indirectly, through their services. Since USDT has become a multi-chain asset, which is also traded and used on other blockchains (e.g., Binance Smart Chain, Avalanche), such shocks could also spread across chains and lead to systemic failures. However, we consider this analysis a first step towards a deeper investigation of systemic risk and keep a deeper investigation for future work.

% !TeX root = ../main.tex

\section{Discussion}\label{sec:discussion}
In this section, we discuss some of the insights from our analyses, as well as the limitations of our work.

\subsection{Insights}

Cryptoassets are not a niche phenomenon anymore. They reached an overall market capitalization of more than 2T USD (Nov. 2021) and are increasingly interconnected with the traditional financial systems.
With DeFi, we now see the introduction of leveraged financial products and assets that are backed with some poorly understood virtual securities.
Our results provide initial insights into the motivating questions mentioned in the introduction.

Concerning ecosystem interoperability, we found that compositions between DEX protocols are particularly frequent in our dataset (c.f. Figure~\ref{plot:fig_pro}). From this, we can conclude that these protocols should ideally be deployed on the same DLT platform as long as single-transaction cross-chain compositions are not possible.
At the same time, however, we also found that derivative protocols in particular still contain relatively few compositions.
This suggests that, for example, a protocol-type specific scaling solution could be useful. For example, a sidechain for derivative protocols. Fewer compositions would still be possible, but not with a significant negative impact as when separating DEX protocols.

As far as integration with web technologies is concerned, the versatile use of building blocks shows that elementary constructs are already reused and integrated by various applications, without this necessarily being transparent to the users.
The view is further reinforced when considering that various assets are already integrated into web technologies, but their simultaneous inclusion in financial instruments and compositions is barely obvious. An example of this is the BAT token, which is integrated into the Brave browser but is also used in various DeFi protocols.

Finally, turning to risks through complexity, we recall that the financial crisis in 2008 has shown that a lack of understanding and lack of regulation can have unforeseeable risks for the financial markets and our society as a whole. 
Whilst composability unleashes unexplored possibilities, it may also lead to unforeseen risks. 
Indeed, despite DeFi protocols are aware of and often even facilitating the use of their own \CAs in composition with those of other protocols, these interconnected novel financial services lack a form of coordination on the resulting compositions. Thus, unintended forms of interaction across protocols could take place, exposing users to risk, even more so when calls are iteratively nested and several protocols are indirectly involved. 
If the DeFi ecosystem evolves at the current pace and integrates closely with the traditional financial sector, associated systemic risks must be understood and mitigated. Our work shows how DeFi protocols can be decomposed, and the share of protocol interactions can be visualized (c.f. Figure~\ref{plot:treemap}). 
With our case study we simulate a hypothetical run on Tether and show how our method can provide first insights how DeFi protocols and their services could be affected, also through cascading effects from other protocols. That shows the potential and possibilities for further studies to evaluate systemic risk.

%

% Limitations
\subsection{Limitations}
\label{sec:limit}
We acknowledge and point out some limitations of our work.
First, our results naturally reflect only the compositions of the protocols and labeled addresses contained in our ground truth dataset. Since the DeFi landscape is evolving rapidly, extending our seed data and the observation period, as well as investigating the temporal evolution of the DeFi protocols, is an obvious next step. One can then re-run our generally applicable analytics procedures. 
We emphasize however that, while a longitudinal analysis of DeFi usage in a longer time frame would be of interest, our main contribution regards the devised methodology to uncover compositions. The time frame and extent of the DeFi protocol activity we investigated are sufficiently large for this (static) analysis. 
Second, as we focused on composability, we didn't investigate some features of the network topology, such as their small-world properties (e.g., clustering coefficients and path lengths); we studied recurrent patterns by decomposing individual transactions as nested building blocks, rather than studying triadic (or higher order) motifs and core decomposition methods; Topological Data Analysis (TDA) has been exploited in the literature mostly in predictive models to identify anomalous patterns, which is beyond the scope of our work; similarly, temporal aspects are left for future work, as discussed previously. In our network analysis, we currently neglect edge weights between \CAs, which may indicate the strength of composition. Including them 
could also be part of future work. 
Third, our building block extraction algorithm currently yields the building blocks of known DeFi protocols. We believe that future work should aim at a more systematic evaluation using a curated ground truth of DeFi compositions.
Finally, we point out that currently we mainly focus on single-transaction interactions between \CAs. However, DeFi compositions could also be constructed by \EOAs over time using multiple transactions. We do not yet consider this aspect in our analysis, but we deem it one of the most promising avenues for future work.

\section{Conclusion}

The overall goal of our work is to provide methods and results that contribute to a better understanding of DeFi protocols, which are a new family of financial products.
% Summary, key take-aways
We manually curated a ground truth set of \DeFiNumberProtocols DeFi protocols, which can be reused in future research. We constructed network abstractions representing the interactions between smart contracts (\CAs) and DeFi protocols and conducted a topology analysis in the timespan from Jan-2021 to Aug-2021. The results indicate the existence of compositions, which is further supported by our finding that known community detection algorithms cannot disentangle DeFi protocols. Therefore, we proposed an algorithm that extracts the building blocks of DeFi protocols from transactions. We assessed the most frequent blocks and found that swaps play an essential role. We also analyzed individual DeFi protocols by disentangling their building blocks and flattened the composition hierarchies of all DeFi protocol transactions in our dataset. We provide a case study, that discovers how the building blocks depend on the USDT stablecoin. This shows how the proposed method can help identify potential systemic risk, by measuring to what extent each protocol is affected by propagating shock of a single entity, originated from vulnerabilities, legal issues or technical advances. Finally, we have discussed the implications and limitations of our work, providing first insights into questions about interoperability, integration with Web technologies, and systemic risks that may arise in complex financial systems. 

In summary, our work is the first that investigates DeFi compositions across multiple protocols, both from a network perspective and at the level of individual transactions.
We believe that our methods make an essential contribution to understanding the bigger picture and the basic building blocks of individual DeFi protocols and their relationships across protocols.

%\begin{acks}
%To Robert, for the bagels and explaining CMYK and color spaces.
%\end{acks}

\bibliographystyle{ACM-Reference-Format}
\bibliography{bibliography/references,bibliography/defi}

%\clearpage % remove this later
%\appendix

%\input{sections/supplemental}
\end{document}